\documentclass[journal]{IEEEtran}
\usepackage{graphicx,amssymb,lineno}
\usepackage{amsmath,amsfonts,amssymb}

\usepackage{algorithm}
\usepackage{algorithmic}
\usepackage[usenames]{color}
\usepackage{float}
\usepackage{stfloats}

\usepackage{mathrsfs}
\usepackage{amsmath}
\definecolor{flexred}{rgb}{0, 0, 0}
\newcommand{\revise}[1]{{\textcolor{flexred}{#1}}}
\definecolor{newred}{rgb}{0.9, 0, 0}

\usepackage{bm}
\usepackage{cite}
\usepackage{url}
\usepackage{hyperref}
\usepackage{multicol}
\usepackage{multirow} 

\usepackage{graphics,color,graphpap,rotate}
\usepackage{times, verbatim, epsfig, latexsym}

\usepackage{graphicx}
\usepackage{subfig}

\begin{document}

\title{Mobility Support for Millimeter Wave Communications: Opportunities and Challenges}

\author{
Jing~Li,~
Yong~Niu,~\IEEEmembership{Member,~IEEE,}
Hao~Wu,~\IEEEmembership{Member,~IEEE,}
Bo~Ai,~\IEEEmembership{{\color{black}Fellow},~IEEE,}
Sheng~Chen,~\IEEEmembership{Fellow,~IEEE,}
Zhiyong~Feng,~\IEEEmembership{Senior Member,~IEEE,}
Zhangdui~Zhong,~\IEEEmembership{{\color{black}Fellow},~IEEE,}
Ning~Wang,~\IEEEmembership{Member,~IEEE,}

\thanks{This study was supported by the National Key Research and Development Program under Grant 2021YFB2900301; in part by the National Natural Science Foundation of China Grants 61801016, 61725101, 61961130391, and U1834210; in part by the National Key R\&D Program of China Grant 2020YFB1806903; in part by the State Key Laboratory of Rail Traffic Control and Safety Grant RCS2021ZT009, Beijing Jiaotong University; and supported by the open research fund of National Mobile Communications Research Laboratory, Southeast University under Grant 2021D09; in part by the Fundamental Research Funds for the Central Universities, China, under Grant 2020JBZD005; and supported by Frontiers Science Center for Smart High-speed Railway System; in part by the Fundamental Research Funds for the Central Universities 2020JBM089; in part by the Project of China Shenhua under Grant GJNY-20-01-1. \emph{(Corresponding author: Yong Niu.)}} %
\thanks{Jing Li, Hao Wu, Bo Ai and Zhangdui Zhong are with the State Key Laboratory of Rail Traffic Control and Safety, Beijing Jiaotong University, Beijing 100044, China, and also with Beijing Engineering Research Center of High-speed Railway Broadband Mobile Communications, Beijing 100044, China (E-mails: jinglee@bjtu.edu.cn,  hwu@bjtu.edu.cn, boai@bjtu.edu.cn, zhdzhong@bjtu.edu.cn).} %
\thanks{Yong Niu is with the State Key Laboratory of Rail Traffic Control and Safety, Beijing Jiaotong University, Beijing 100044, China, and also with the National Mobile Communications Research Laboratory, Southeast University, Nanjing
211189, China (E-mail: niuy11@163.com).}
\thanks{Sheng Chen is with the School of Electronics and Computer Science, University of Southampton, Southampton SO17 1BJ, U.K. (E-mail: sqc@ecs.soton.ac.uk).}
\thanks{Zhiyong Feng is with Beijing University of Posts and Telecommunications, Key Laboratory of Universal Wireless Communications, Ministry of Education, Beijing 100876, P. R. China (E-mail:
fengzy@bupt.edu.cn).}
\thanks{Ning Wang is with the School of Information Engineering, Zhengzhou University, Zhengzhou 450001, China (E-mail: ienwang@zzu.edu.cn).}
\vspace*{-5mm}
}

\maketitle

\begin{abstract}
Millimeter-wave (mmWave) communication technology offers a potential and promising solution to support 5G and B5G wireless networks in dynamic scenarios and applications. However, mobility introduces many challenges as well as opportunities to mmWave applications. To address these problems, we conduct a survey of the opportunities and technologies to support mmWave communications in mobile scenarios. Firstly, we summarize the mobile scenarios where mmWave communications are exploited, including indoor wireless local area network (WLAN) or wireless personal area network (WPAN), cellular access, vehicle-to-everything (V2X), high speed train (HST), unmanned aerial vehicle (UAV), and the new space-air-ground-sea communication scenarios. Then, to address users' mobility impact on the system performance in different application scenarios, we introduce several representative mobility models in mmWave systems, including human mobility, vehicular mobility, high speed train mobility and ship mobility. Next we survey the key challenges and existing solutions to mmWave applications, such as channel modeling, channel estimation, anti-blockage, and capacity improvement. Lastly, we discuss the open issues concerning mobility-aware mmWave communications that deserve further investigation. In particular, we highlight future heterogeneous mobile networks, dynamic resource management, artificial intelligence (AI) for mobility and integration of geographical information, deployment of large intelligent surface and reconfigurable antenna technology, and finally, the evolution to Terahertz (THz) communications.
\end{abstract}


\section{Introduction}\label{S1}

With the various emerging applications in the 5G era, extensive mobile connections occur in all walks of life, from technological to social activities, which exhibit two major trends. Firstly, rapid traffic growth appears in large cities or social life hotspots that are highly related to humans' movement. It is predicted that by 2022 the world mobile data traffic will reach 77 ExaBytes per month, exceeding six times that in 2017 \cite{Cisco2019}. This capacity demand introduces a heavy burden on the limited spectrum at sub-6\,GHz. Secondly, the construction of intelligent transportation system (ITS) highly relies on massive mobile connections among vehicle to vehicle, railway to infrastructure, and road station to passengers. 
{\color{black}Some promising applications, such as automated driving, impose extra concerns on low latency and high reliability in mobile communications.}
{\color{black}Plentiful spectrum resources} are needed to fulfill these transmission demands of massive mobile traffic at low latency in these dynamic communication scenarios as well as emerging new applications \cite{Intro2018}. Millimeter-wave (mmWave) communications, which tap into large available bandwidth resources of the mmWave band, are witnessed to be a potential solution to support these trends.

\revise{Related research efforts have been made on mmWave technology since 1990s.}
Initially, rapid progress emerged in complementary metal-oxide-semiconductor (CMOS) radio frequency (RF) integrated circuits, paving the way for mmWave viability. {\color{black}For example, the fully CMOS-based beamforming receiver at 60\,GHz achieves high performance at low cost and has become popular in commercial applications \cite{CMOS}.} In 2013, Rappaport {et al.} \cite{mCA1} conducted mmWave (28\,GHz, 38\,GHz) channel measurement in urban environments, making a valuable step towards the 5G cellular communications at mmWave bands. {\color{black}Since then, extensive measurement campaigns in mmWave frequency bands (41\,GHz, 60\,GHz, 73\,GHz, etc.) have been conducted to characterize propagation channels.} 
\revise{MmWave wireless communication suffers from high path loss and must rely heavily on directional large-scale
antenna array for power concentration. This imposes a serious challenge on how to provide reliable mmWave connections for highly dynamic scenarios. However, there exists limited progress on mobility support for mmWave communications until the deployment of adaptive beamforming techniques. Two advances are incorporated with such techniques, directional transmission with beamforming compensating for the propagation loss of mmWave signals and robust adaptation to the fast-changing environment for enabling beam alignment in dynamic scenarios.}

Specifically, smart motion-prediction beam alignment (SAMBA) \cite{V2X-2021} and fast machine learning (FML) algorithm \cite{V2XINF} have recently been adopted in mmWave vehicle-to-everything (V2X) communications. Furthermore, researchers have exploited enabling technologies to improve the mmWave network performance in dynamic scenarios, including combining it with massive multiple-input and multiple-output (MIMO), device-to-device (D2D) communications, relaying, new spatial processing techniques, mobile management techniques, etc.  \cite{RelayJ1,NiuD2D,NiuMIMO,mWadd1,mWadd4,mWadd6,mWadd7,CATWC,mCA_10}. For instance, \cite{CATWC} and \cite{mCA_10} prompted coverage optimization by integrating MIMO and D2D in mmWave systems, respectively. To date, there are already standards available for wireless networks (WLAN) and wireless personal area networks (WPAN) at mmWave frequency, such as IEEE 802.11ad and IEEE 802.15.3c, while the standardization activities for mmWave V2X networks, such as 802.11bd, are currently being promoted by {\color{black}IEEE} \cite{NiuSurvey,Standrad1,11bd,V2X-2021a}.
                   
\revise{Several surveys \cite{XiaoMing,11bd,NiuSurvey,SurveyBAi,SurveyM} have summarized {\color{black}the progress} in mmWave communications and provided insightful understanding on the development of mmWave networks. {\color{black}Yet the important research for mobility support is still ongoing, which is facing the following challenges:}  
\begin{itemize}
\item {\color{black}Diverse scenarios}: Mobility scenarios vary from low-speed WLANs to high-speed railway networks, which serve distinct applications and impose different requirements for interference management, beamforming design, energy efficiency, etc. Therefore, a categorized review of mmWave communications research conducted {\color{black}in scenarios} with different mobility levels is necessary. 
\item {\color{black}Mobility modeling}: The goal of mobility modeling is to imitate real-life mobility based on the extracted characteristics in the underlying scenarios. Due to the complexity of mobility data collection, filtering, and fusion in heterogeneous networks (HetNets), there are few works on mobility feature analysis in the mmWave spectrum.
\item {\color{black}Related critical issues}: Mobility introduces many problems in mmWave networks, including inaccurate channel measurement, complex channel modeling, dynamic channel estimation, frequent blockage, capacity decrease, etc. Yet a comprehensive work is still lacking in analyzing these problems and discussing the existing solutions. How to integrate the innovative techniques into the upcoming mobile mmWave networks and further enhance the system performance have not been fully considered.
\end{itemize}
}
\begin{figure}[!htp]
\vspace*{-2mm}
\begin{center}
\includegraphics[width=\columnwidth]{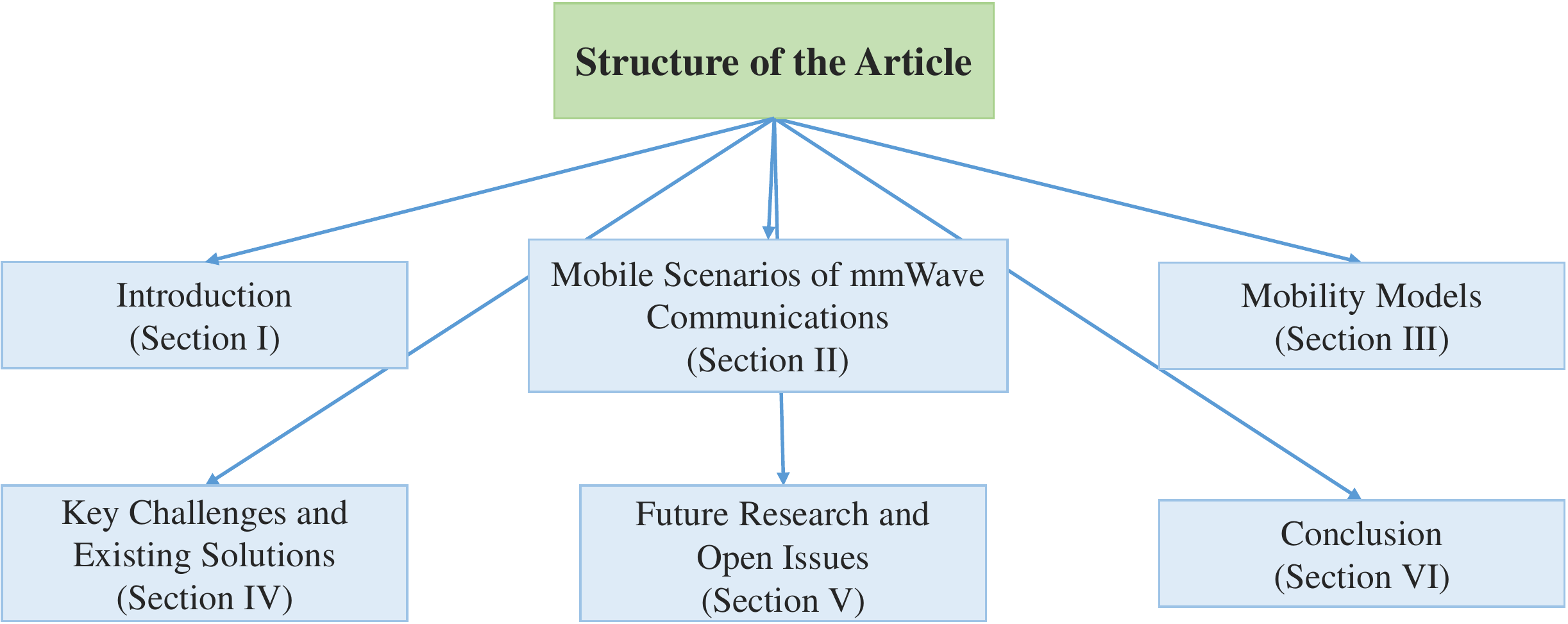}
\end{center}
\vspace*{-2mm}
\caption{\revise{Structure of the survey article.}}
\label{fig:Introduction} 
\vspace*{-2mm}
\end{figure}

Motivated by these problems, we carry out a survey to investigate the challenges and technologies to support mmWave communications in mobile scenarios. 
To serve this purpose, a deep insight {\color{black}is provided} for the influence of mobility in mmWave networks, and the existing solutions {\color{black}to} key challenges are summarized comprehensively. The structure of this paper is shown in Fig.~\ref{fig:Introduction}. 
First, we introduce the mobile scenarios where mmWave communications are exploited, in Section~\ref{S2}. In different mobile scenarios, mobility patterns are different. Thus, we next introduce different mobility models to be considered in the design of mmWave systems in Section~\ref{S3}. Then in Section~\ref{S4}, we carry out an extensive literature review on mmWave communications and emphasize the mobility support for mmWave communications in mobile scenarios. The challenges from the physical layer to the network layer to support mobility are also discussed in this section. Furthermore, we list the open research issues for mmWave communications to support mobility in mobile scenarios in Section~\ref{S5}. Finally, we conclude this survey paper in Section~\ref{S6}.

\section{Applications of mmWave Communications in Mobile Scenarios}\label{S2}

{\color{black}By leveraging the abundant bandwidth and directional beamforming, mmWave communications can be applied to a host of mobile scenarios\cite{UAV_zhenyuxiao}. However, serving different applications in these scenarios brings various challenges to the design of communication systems. To overcome these challenges, extensive research has been conducted in channel modeling, beamforming design, energy efficiency, standards development, etc. The key technologies, potential solutions and existing difficulties in different scenarios are summarized below.}

\subsection{Indoor WLAN and WPAN}\label{S2.1}

IEEE\,802.15.3c and IEEE\,802.11ad define the standards for indoor WLANs and WPANs, respectively, in the 60\,GHz band \cite{IEEE_802.15.3c,IEEE_802.11ad}, directly enabling the application of mmWave communications in indoor WLAN and WPAN. With huge unlicensed bandwidth in the 60\,GHz band, multi-gigabit per second (Gbps) {\color{black}transmission rates} can be achieved, and wideband multimedia applications, including high-rate data transfer between devices, {\color{black}such as cameras, pads, personal computers, etc., are available}.

Research on this area is continuing. For example, the work \cite{mWwlan2012} investigated mmWave wireless link performance in indoor environments. Che {et al.} \cite{mWwlan2014} carried out a joint design of axis alignment and positioning for non-line-of-sight (NLoS) indoor mmWave WLAN and WPANs. A link scheduling algorithm was proposed for mmWave WPAN in \cite{mWwlan2016}, which includes common channel interference probing scheme, link coexistence determination scheme and link schedule algorithm, to enhance the performance. The paper \cite{mWwlan2021} investigated access point (AP) placement for multi-AP mmWave WLANs. {\color{black}In summary, mmWave systems coexist well with indoor WLAN and WPAN systems. Table~\ref{table-WLAN} summarizes the overall research discussed in this subsection.} %

\begin{table}[hb!]
\vspace*{-4mm}
\caption{{\color{black}Research on mmWave Indoor WLAN/WPAN}}
\label{table-WLAN} 
\vspace*{-4mm}
\begin{center}
\resizebox{0.5\textwidth}{!}{
\begin{tabular}{|l|l|l|l|} 
\hline
Ref. & Year & {\color{black}Technique/Contribution} & {\color{black}Utilization/Application}\\ \hline
\cite{mWwlan2012} & 2012 &\begin{tabular}[c]{@{}l@{}}MmWave propagation\\ prediction\end{tabular} &{\color{black}Performance evaluation}\\ \hline
\cite{mWwlan2014} & 2014 & \begin{tabular}[c]{@{}l@{}} Joint design of axis\\ Alignment and positioning  \end{tabular}&{\color{black}Performance improvement}
\\ \hline
\cite{mWwlan2016} & 2016 & Link scheduling &{\color{black}Capacity improvement} \\ \hline
\cite{mWwlan2021} & 2021 & Multi-AP placement  &{\color{black}\begin{tabular}[c]{@{}l@{}}Line-of-sight coverage \\ maximization\end{tabular}}
\\ \hline
\end{tabular}
}
\end{center}
\vspace*{-5mm}
\end{table}

\subsection{Cellular Access}\label{cellular access app} 

MmWave communication technologies have long attracted attention as a promising solution for {\color{black}next-generation} mobile communications. But not until 2013 that researchers have found that mmWave may be suitable for cellular networks and has advantages over microwave-based cellular networks \cite{mCA1}. However, as mmWave communication has different characteristics from its microwave counterpart, the currently available sub-6\,GHz frameworks cannot be applied directly. Hence recent works have extensively studied the application of mmWave technology in cellular access.

In 2011, Andrews et al. \cite{mCA_2} put forward an analytical approach for analyzing coverage and rate of classic cellular networks. Since then considerable progress has been made to modify the model, validate its accuracy and extend its applications to more sophisticated mobile network scenarios \cite{cnA3,cnA4,cnA5}. An analytical framework for mmWave cellular networks was studied in \cite{mCA_3,mCA_6}. Specifically, based on the theory of Poisson point process (PPP), Renzo \cite{mCA_3} {\color{black}leveraged stochastic geometry} to build the analytical framework. With realistic channel and blockage models for mmWave propagation, sufficiently dense mmWave cellular networks were shown to outperform microwave cellular networks. Rebato et al. \cite{mCA_6} further applied an experimental mmWave channel model to this framework and derived the signal-to-interference ratio (SIR) based coverage probability in mmWave cellular networks. Moreover, the accuracy of the PPP-based framework was studied in \cite{mCA_4}. By taking realistic base station (BS) locations, buildings footprints, and empirical mmWave channel models into account, it was shown that this framework is capable of estimating the downlink performance in dense mmWave cellular networks. The framework was further improved by considering other-cell interference in \cite{mCA_5}, and Monte Carlo simulations were conducted to validate its performance in computing the coverage of mmWave cellular networks. {\color{black}In practice, however, this framework may not be applicable for D2D communications and cognitive networks, where users with the same demands tend to form a cluster, but the PPP-based model fails to describe this property accurately. Therefore, Poisson cluster process (PCP) theory considering multiple clusters is leveraged in related research \cite{ClusterTCOM}.}

On the other hand, several system models are proposed to overcome specific challenges in mmWave cellular networks \cite{mCA_7,mCA_11,mCA_8,mCA_9,mCA_10}. Considering the high propagation path loss and sensitivity to blockages in mmWave band, Elkotby and Vu \cite{mCA_7} employed MIMO beamforming in mmWave cellular networks, and proposed a probabilistic interference distribution model, in which line-of-sight (LoS) interference power is characterized as a Gamma distribution, while NLoS interference power is modeled as a mixture of the inverse Gaussian and inverse Weibull distributions. Petrov et al. \cite{mCA_11} constructed a novel methodology based on queuing theory and stochastic geometry to solve the problems of complex radio propagation, human mobility, and multi-connectivity in mmWave cellular networks. Singh et al. \cite{mCA_8} developed a tractable model to capture the user rate distribution and to derive the rate expression in mmWave cellular networks. With this developed model and based on simulations, it was shown that mmWave cellular networks are noise-limited and the rate heavily relies on the BS density. Besides, Mezzavilla et al. \cite{mCA_9} proposed a Markov decision process (MDP) model to study handover problems in mmWave cellular networks. Umer et al. \cite{mCA_10} analyzed the performance of mmWave cellular networks coexisting with microwave cells, and found that massive MIMO and densely deployed mmWave cells significantly enhance rate and coverage.
Zhou et al. \cite{mmWave2018} developed a hardware-efficient hybrid precoding scheme for mmWave systems based on novel multi-feed reflect arrays. {\color{black}Based on graph theory, Sha et al. \cite{Cell2020} proposed a time-domain beam scheduling for the mmWave cellular network.}
{\color{black}To give a quick view of these efforts, we summarize significant progress on mmWave cellular networks in Table~\ref{table-Cellular}.}

\begin{table}[tp!]
{\color{black}
\caption{{\color{black}Research on mmWave Cellular Access }}
\label{table-Cellular} 
\vspace*{-4mm}
\begin{center}
\resizebox{0.5\textwidth}{!}{
\begin{tabular}{|l|l|l|l|} 
\hline
Ref. & Year &{\color{black}Technique/Contribution}& {\color{black}Utilization/Application} \\ \hline
\cite{mCA1}   & 2013 &\begin{tabular}[c]{@{}l@{}}Survey of mmWave 5G \\cellular system\end{tabular}&{\color{black}-}  \\ \hline
\cite{mCA_8}  & 2014 &Key features analysis & {\color{black}User rate improvement} \\ \hline
\cite{mCA_3}  & 2015 &\begin{tabular}[c]{@{}l@{}}Analytical framework\\ proposition\end{tabular} &{\color{black}\begin{tabular}[c]{@{}l@{}}Coverage and rate\\ analysis \end{tabular}} \\ \hline
\cite{mCA_4}  & 2015 &Network modeling & {\color{black}\begin{tabular}[c]{@{}l@{}}Coverage and rate\\ analysis \end{tabular}}\\ \hline
\cite{mCA_5}  & 2015 &\begin{tabular}[c]{@{}l@{}}PPP-based framework\\ improvement\end{tabular}& {\color{black}Coverage analysis} \\ \hline
\cite{mCA_7}  & 2016 &MIMO beamforming &{\color{black}Interference modeling} \\ \hline
\cite{mCA_9}  & 2016 &Handover management &{\color{black}Capacity improvement} \\ \hline
\cite{mCA_6}  & 2017 &Analytical model proposition&{\color{black}Coverage analysis} \\ \hline
\cite{ClusterTCOM} &{\color{black}2017} &{\color{black}\begin{tabular}[c]{@{}l@{}}Network modeling based on\\ PCP\end{tabular}} &{\color{black}Coverage analysis}\\ \hline
\cite{mCA_11} & 2017 &Multi-connectivity &{\color{black}\begin{tabular}[c]{@{}l@{}}Session continuity \\improvement\end{tabular}} \\ \hline 
\cite{mmWave2018} & 2018 &Hybrid precoding&  - \\ \hline
\cite{Cell2020} &{\color{black}2020} & {\color{black}Beam scheduling} &{\color{black}Interference avoidance}
\\ \hline
\end{tabular}
}
\end{center}
}
\vspace*{-5mm}
\end{table}

\subsection{V2X Communications}\label{S2.3}

V2X communications demand high data rates to support more advanced vehicular applications, such as automated driving and in-vehicle infotainment systems. It is estimated that transmission rates in the order of gigabits per driving hour are needed \cite{V2X_1}. However, existing technologies cannot satisfy this demand. Current achievable data rates for vehicle-to-vehicle (V2V) communications and vehicle-to-infrastructure (V2I) communications are respectively 2-6 Mbps and 100 Mbps \cite{V2X_1}. Therefore, the next-generation mmWave technology is suggested in V2X communications for rate improvement. MmWave-based V2X communications face the usual challenges, such as high propagation loss and sensitivity to blockage. Moreover, mmWave V2X communications face extra serious challenges {\color{black}in that} stability, security and ultra-low latency are necessary for traffic purposes. 
\revise{Numerically, \cite{{V2XLatency}} specified the maximum communication latencies for automated overtake and high-density platooning as 10\,ms.}

There exist many works focusing on the mmWave application to V2X communications. Perfecto et al. \cite{V2X_2} proposed a distributed multi-beam association scheme for mmWave vehicular scenarios to expand the individual sensing range of vehicles. This proposed scheme is capable of increasing the average volume of collected sensed information by up to 71\%.
\revise{Rudraksh et al. \cite{V2X_3} developed a vehicular version of the shared user equipment-side (UE-side) distributed antenna system (SUDAS), which leverages the UE-side radio units (URUs) on the vehicle body to transform the outdoor MIMO signal into a single-input and single-output (SISO) signal on the mmWave channel and transmits it to the UEs. Such a strategy enables high throughput and reliable wireless communication in high mobility V2X scenarios.}

The work \cite{V2X-2018} designed a novel redundancy-based preamble transmission in order to fast acquire data transmission opportunities in mmWave-based massive V2X communications scenarios. Brambilla et al. \cite{V2X-2020} developed a location-assisted and subspace-based beam alignment scheme for LoS and NLoS mmWave V2X Communications. The work \cite{V2X-2021} proposed the novel intelligent beam control and secure stable routing scheme to address beam alignment difficulties and routing stability issues due to rapid mobility of vehicles, in order to support mmWave-based ultra-low-delay V2X transmissions. The paper \cite{V2X-2021a} provided a detailed assessment on the capability of mmWave V2X to support automated driving, discussed specific challenges related to mmWave for V2X, and \revise{pointed out the mmWave V2X standards being developed by IEEE and 3GPP, IEEE 802.11bd and 3GPP new radio (NR) V2X.}
{\color{black}For clarity, Table~\ref{table-V2X} summarizes the literature review for mmWave V2X applications.}

\begin{table}[h!]
\vspace*{-1mm}
{\color{black}
\caption{{\color{black}Research on mmWave V2X Applications}}
\label{table-V2X} 
\vspace*{-4mm}
\begin{center}
\resizebox{0.5\textwidth}{!}{
\begin{tabular}{|l|l|l|l|} 
\hline
Ref. & Year &{\color{black}Technique/Contribution} & {\color{black}Application} \\ \hline
\cite{V2X_2}     & 2017 & \begin{tabular}[c]{@{}l@{}}Distributed multi-beam\\ Association\end{tabular} &{\color{black}\begin{tabular}[c]{@{}l@{}}Sensing range expansion\end{tabular}}\\ \hline
\cite{V2X_3}     & 2017 & SUDAS development &{\color{black}\begin{tabular}[c]{@{}l@{}}Throughput and reliability\\ enhancement\end{tabular}}\\ \hline
\cite{V2X-2018}  & 2018 & \begin{tabular}[c]{@{}l@{}}Fast acquisition of \\transmission opportunities\end{tabular} &{\color{black}\begin{tabular}[c]{@{}l@{}} Access reliability\\improvement\end{tabular}}\\ \hline
\cite{V2X-2020}  & 2020 & \begin{tabular}[c]{@{}l@{}}Location-assisted and\\ subspace-based beam\\ alignment\end{tabular} &{\color{black}Performance improvement}\\ \hline
\cite{V2X-2021}  & 2021 & Intelligent beam control &{\color{black}Performance improvement}\\ \hline
\cite{V2X-2021a} & 2021 & Capability assessment &{\color{black}\begin{tabular}[c]{@{}l@{}}Automated driving support\end{tabular}}\\ \hline
\end{tabular}
}
\end{center}
}
\vspace*{-6mm}
\end{table}

\begin{figure}[bp!]
\vspace*{-6mm}
\begin{center}
   \subfloat[Tunnel]{\includegraphics[height=2.8cm]{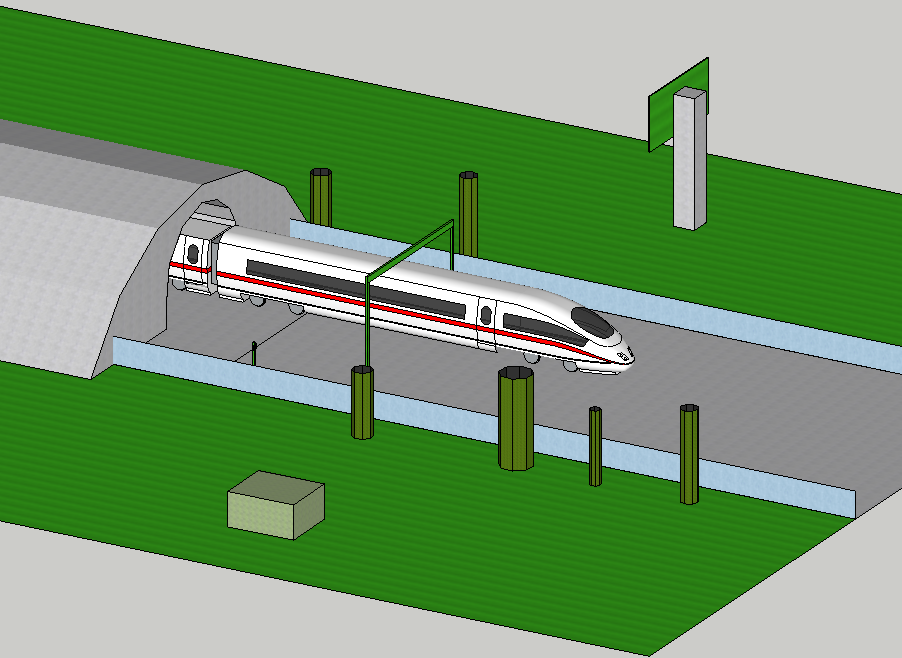}}~~
  \subfloat[Urban]{\includegraphics[height=2.8cm]{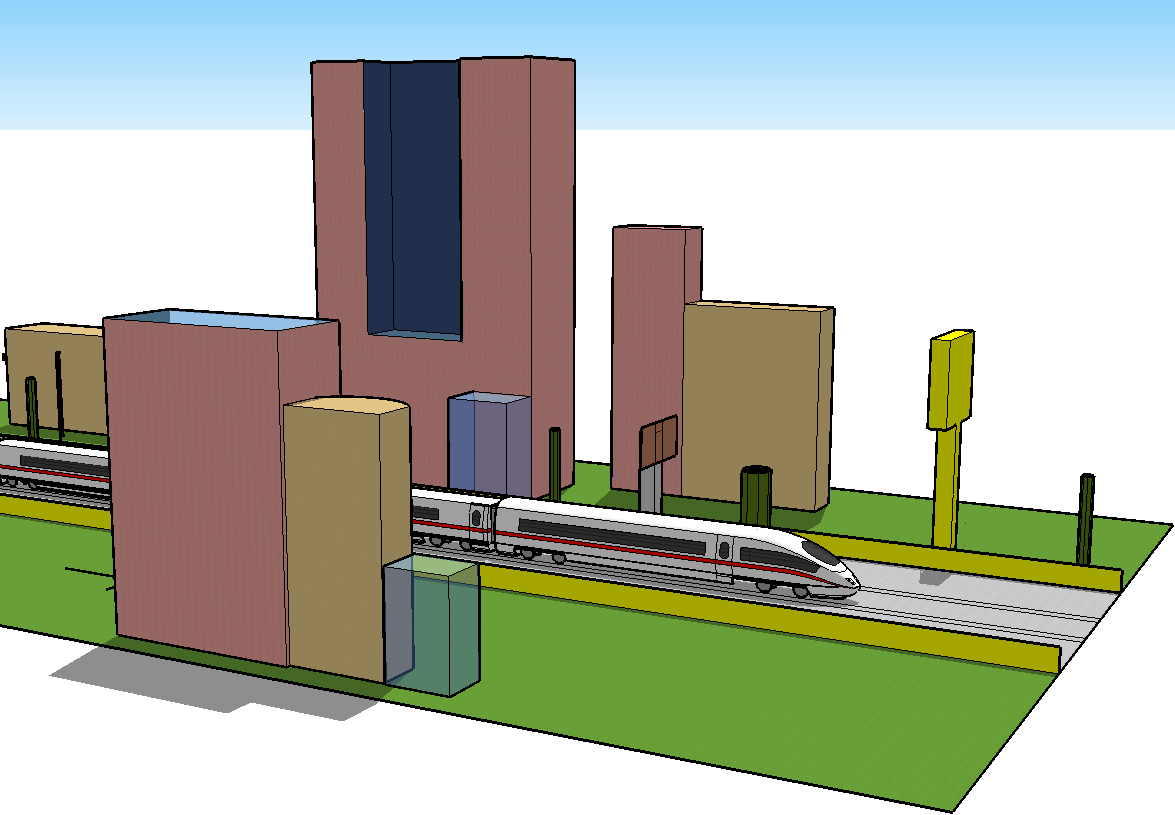}}\
  \
  \subfloat[Cutting]{\includegraphics[height=2.8cm]{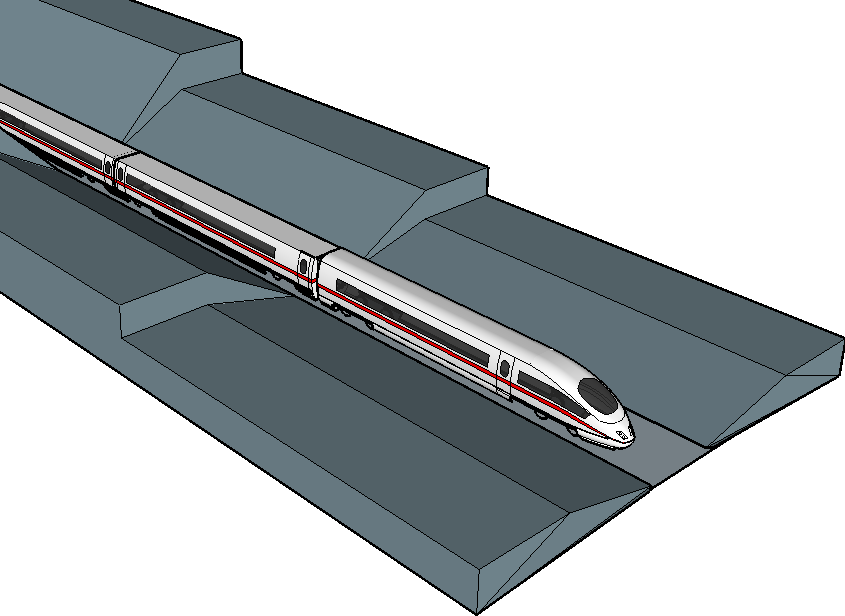}}~~
  \subfloat[Viaduct]{\includegraphics[height=2.8cm]{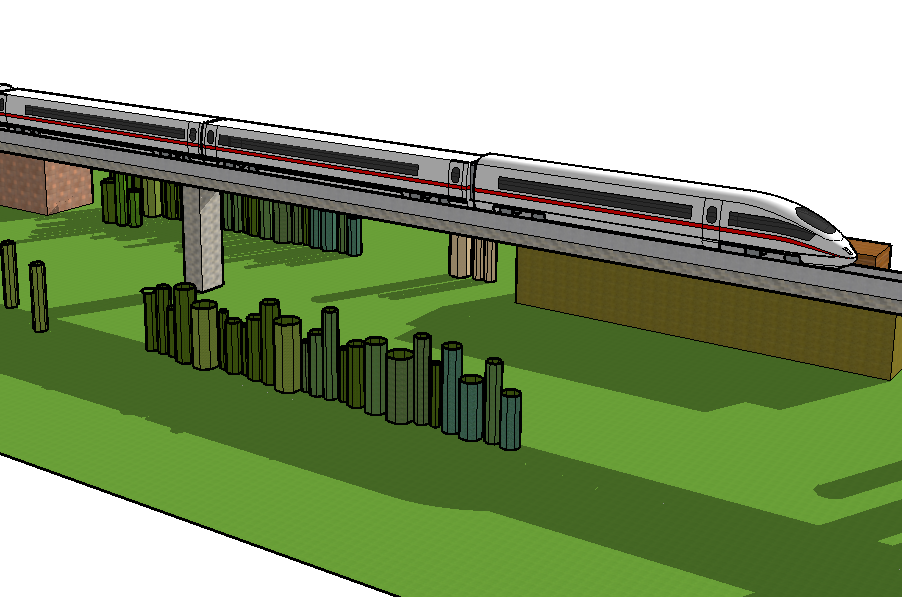}}
\end{center}
\vspace*{-3mm}
\caption{\revise{Examples of complex environments in HSTs.}}
\label{fig:complex} 
\vspace*{-1mm}
\end{figure}

\subsection{High Speed Train}\label{S2.4}

In high speed train (HST) scenarios, communication conditions need improvement urgently. On the one hand, HST is becoming faster and faster (350\,km/h and above), requiring even faster disaster detection systems and imposing increasingly challenging communication environments, \revise{especially when going through complex sections illustrated by Fig. 2.} On the other hand, customers aboard HST are calling for better online experiences. \revise{These requirements prompt a shift to intelligent mobility management and demand high capacity and high data rate wireless transmission in the future HST system.}
 
\revise{Accordingly, growing attention has been focused on mmWave technology owing to its wide bandwidth and high communication rates.}
In order to overcome the high path loss, however, mmWave-based HST communications require frequent realignment and beamforming technology.
{\color{black}Extensive research has been} carried out in the past decade.
Kim et al. \cite{HST_1} designed a beamforming scheme based on beam pattern, and provided an insight of applying beamforming in mmWave-based HST communications. He et al. \cite{HST_4} proposed a beamforming scheme to improve the disaster detection efficiency and decreased the false alarm rate in HST scenarios. Cui et al. \cite{HST_2} presented a hybrid spatial modulation scheme and validated its performance for HST communications. On the other hand, conventional beam sweeping schemes and wide beams become inefficient due to the short coherence time in mmWave-based HST communications. Thus, Va et al. \cite{HST_5} came up with a beam switching approach \revise{and investigated the optimal choice of beamwidth, with which the mmWave system can achieve multi-Gbps throughput in HSTs.}
Kim et al. \cite{HST_6} proposed a distributed antenna system for mmWave mobile broadband communications, and showed that it is possible to provide Gbps data services.

{\color{black}More recently, the research \cite{HST-2017,HST-2020TAP} specifically investigated mmWave channel characteristics for HST in tunnel and intra wagon scenarios.} {\color{black}The work \cite{HST-2019} presented a design of the mmWave system dedicated to HST communications between train and trackside.} In this paper, a frame structure was proposed for the acquirement of channel state information (CSI) and the transmission of user data. Moreover, the channel measurement results obtained under high-speed mobile conditions were provided, and the throughput of the system prototype recorded during the transmission of three high-definition video streams {\color{black}was} presented. Cheng et al. \cite{HST-2019a} proposed a fast beam searching scheme to reduce the number of measurements required for mmWave {\color{black}communications} in HSTs.
{\color{black}For clarity, Table~\ref{table-HST} lists the research on mmWave HST applications. In summary, mature channel models for mmWave HST communications have not been developed yet, and beam alignments and performance maintenance in such a high-speed scenario are still challenging.} 

\begin{table}[h!]
\vspace*{-1mm}
{\color{black}
\caption{{\color{black}Research on mmWave HST Applications}}
\label{table-HST} 
\vspace*{-4mm}
\begin{center}
\resizebox{0.5\textwidth}{!}{
\begin{tabular}{|l|l|l|l|} 
\hline
Ref. & Year &{\color{black}Technique/Contribution}& {\color{black}Application} \\ \hline
\cite{HST_6}     & 2013 & \begin{tabular}[c]{@{}l@{}}Distributed antenna\\ system \end{tabular} &{\color{black}\begin{tabular}[c]{@{}l@{}}Supporting high data rate \end{tabular}}\\ \hline
\cite{HST_1}     & 2015 & Beamforming design &{\color{black}Guiding system design}\\ \hline
\cite{HST_5}     & 2015 & Beam switching &{\color{black}\begin{tabular}[c]{@{}l@{}}Throughput improvement\end{tabular}}\\ \hline
\cite{HST_4}     & 2016 & Beamforming design &{\color{black}Disaster detection}\\ \hline
\cite{HST_2}     & 2016 & Hybrid spatial modulation &{\color{black}\begin{tabular}[c]{@{}l@{}}Performance enhancement\end{tabular}}\\ \hline
\cite{HST-2017}  & 2017 & \begin{tabular}[c]{@{}l@{}}Investigating channel\\  characteristics in tunnel\end{tabular} & {\color{black}\begin{tabular}[c]{@{}l@{}}Reliability enhancement\end{tabular}}\\ \hline
\cite{HST-2019}  & 2019 & \begin{tabular}[c]{@{}l@{}}System design between\\ train and communication\end{tabular} &{\color{black}\begin{tabular}[c]{@{}l@{}}Robustness improvement\\ and spectrum efficiency\\ enhancement\end{tabular}}\\ \hline
\cite{HST-2019a} & 2019 & Fast beam searching &{\color{black}\begin{tabular}[c]{@{}l@{}}Reducing measurement\\ requirements\end{tabular}}\\ \hline
{\color{black}\cite{HST-2020TAP}} &{\color{black}2020} &{\color{black}\begin{tabular}[c]{@{}l@{}}Channel sounding for \\intra wagon\end{tabular}} &{\color{black}\begin{tabular}[c]{@{}l@{}}Channel characterization\end{tabular}}\\ \hline
\end{tabular}
}
\end{center}
}
\vspace*{-7mm}
\end{table}

\subsection{Unmanned Aerial Vehicle}\label{S2.5}

Unmanned aerial vehicles (UAVs) are widely used both in military and commercial fields, and they have become prevalent in our daily life. Under many circumstances, like reconnaissance, remote sensing and aerial photography, a large amount of data from various sensors need to be sent back to control stations as fast as possible. Therefore, high data rates are of great significance in UAV communications. Owing to fast deployment and flexible reconfiguration, 
 {\color{black}UAV-aided} communications can be exploited to enhance the capacity and services of the existing cellular network. They are also particularly useful to provide broadband services to the remote part of the world, where communications infrastructures do not exist. More specifically, UAVs can be employed as different types of wireless communication platforms, such as UAV base stations (UAV-BS), aerial relays, and UAV swarms \cite{UAV-2018}.

{\color{black}Compared to terrestrial mmWave communications, the propagation characteristics in mmWave UAV communications are very unique because of the 3D blockage, aircraft shadowing, and UAV fluctuation \cite{NewUAV1}. 
In addition, traditional beam tracking methods are deficient to predict the beams with UAVs’ rotation in 3D space \cite{VVa, UAVICCIr}.}
Thus, besides common challenges of range and directional communications, special issues, like {\color{black}accurate channel characterization}, fast channel tracking, more efficient beamforming training, accurate trajectory prediction and loading capacity, should be taken into consideration{\color{black}\cite{UAVChannelModeling}}, \cite{UAV_1}. Recently, considerable research efforts have been directed towards UAV-based mmWave communications. Using the channel model incorporated with the distance-based random blockage effects, which is based on stochastic geometry and random shape theory, Jung and Lee \cite{UAV-2017} investigated the outage performance of the mmWave UAV swarm network. The authors \revise{also} showed how to minimize the outage rate by adjusting various system parameters. Zhang et al. \cite{UAV-2019} surveyed key technical advantages, challenges and potential applications for UAV-assisted mmWave networks. The authors of \cite{UAV-2019a} developed a 3D beamforming approach to achieve efficient and flexible coverage in mmWave UAV-BS communications. The work \cite{UAV-2020} proposed an empirical propagation loss model for UAV-to-UAV communications at mmWave band, based on an extensive aerial measurement campaign conducted with the Facebook Terragraph channel sounders.

{\color{black}For a quick view of the above work, we summarize the research on mmWave UAV applications in Table~\ref{table-UAV}. Future designs could concentrate more on clustered mmWave UAV networks, where the research on channel modeling, beam switching and coverage analysis are in its initial phase.}

\begin{table}[h!]
{\color{black}
\caption{{\color{black}Research on mmWave UAV Applications}}
\label{table-UAV} 
\vspace*{-4mm}
\begin{center}
\resizebox{0.5\textwidth}{!}{
\begin{tabular}{|l|l|l|l|} 
\hline
Ref. & Year &{\color{black}Technique/Contribution}& {\color{black}Utilization/Application} \\ \hline
\cite{UAV_1}     & 2016 & \begin{tabular}[c]{@{}l@{}}Survey of mmWave UAV\\ systems\end{tabular} &{\color{black}-}\\ \hline
\cite{UAV-2017}  & 2017 & Outage performance analysis & {\color{black}Coverage enhancement}\\ \hline
\cite{UAV-2018}  & 2018 & \begin{tabular}[c]{@{}l@{}}Survey of mmWave UAV \\systems\end{tabular} &{\color{black}-}\\ \hline
\cite{NewUAV1}   & 2019 & \begin{tabular}[c]{@{}l@{}}Survey of mmWave UAV \\systems\end{tabular} &{\color{black}-}\\ \hline
\cite{UAV-2019}  & 2019 & \begin{tabular}[c]{@{}l@{}}Survey of mmWave UAV for\\ 5G\end{tabular} &{\color{black}-}\\ \hline
\cite{UAV-2019a} & 2019 & 3D beamforming  &{\color{black}Coverage enhancement}\\ \hline
\cite{UAV-2020}  & 2020 &\begin{tabular}[c]{@{}l@{}} Empirical propagation model  \end{tabular} &{\color{black}Performance analysis}\\ \hline
\end{tabular}
}
\end{center}
}
\vspace*{-6mm}
\end{table}

\begin{figure}[!bp]
\vspace*{-5mm}
\begin{center}
\includegraphics[width=\columnwidth]{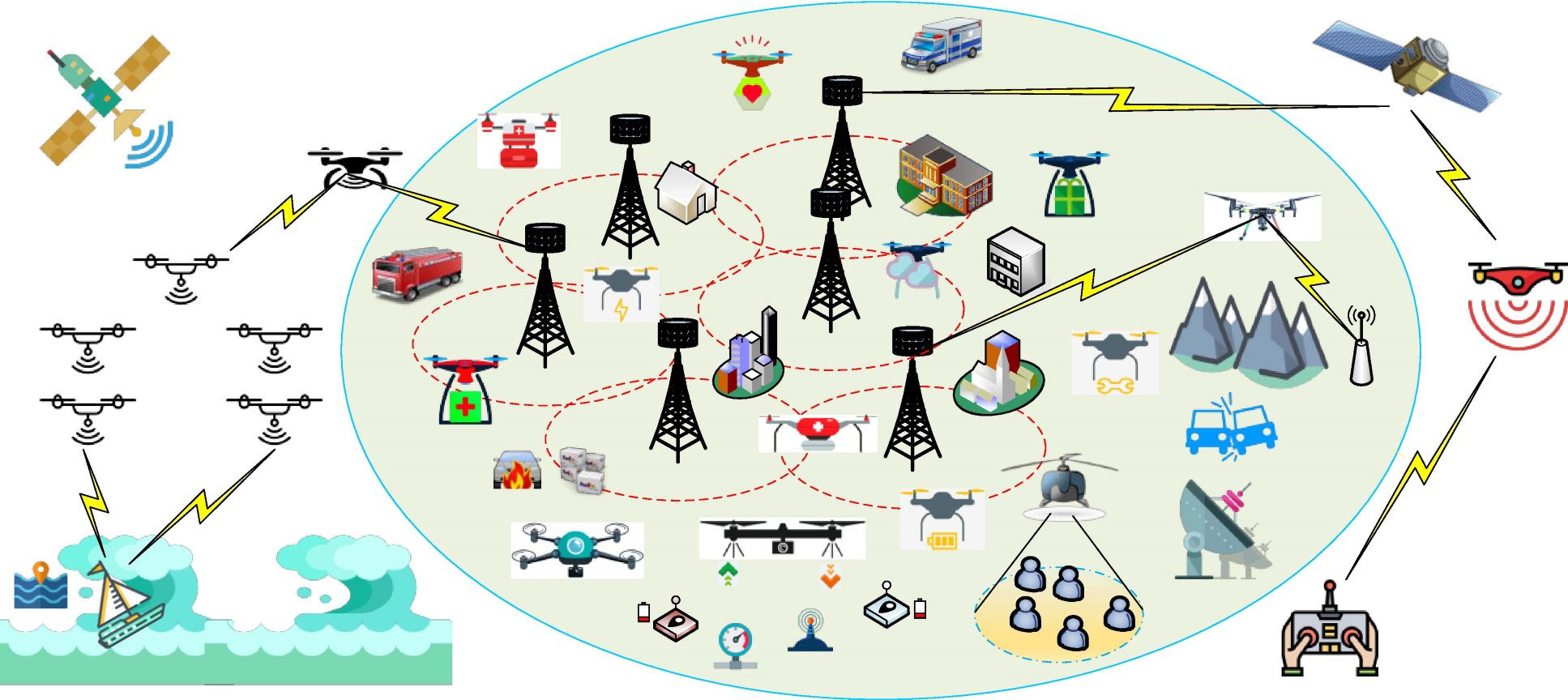}
\end{center}
\vspace*{-2mm}
\caption{The space-air-ground-sea network \cite{UAV1}.}
\label{fig:UAV} 
\vspace*{-1mm}
\end{figure}

\subsection{Space-Air-Ground-Sea}\label{S2.6}

Even when 5G implementation is only started in few countries, researchers are already thinking the {\color{black}next-generation} mobile network. To extend the terrestrial 5G to cover every part of the world, Chen \cite{6G-2019} proposed the concepts of aeronautical ad hoc network (AANET) to realize Internet above the cloud and oceanic ad hoc network (OANET) to realize Internet above the wave. Although 5G is still in its infancy with most people around the world are still on 4G, the race towards 6G has already started. On November 6, 2020, China launched the world's first 6G test satellite into orbit to verify the Terahertz (THz) communication technology in space \cite{6G-news}. As illustrated in Fig.~\ref{fig:UAV}, the future 6G will be the space-air-ground-\revise{sea} (SAGS) network, combining satellite, air, sea and terrestrial communications to offer seamless coverage and stable broadband services for users any where and any time. {\color{black}This grand SAGS network will integrate the world's satellite networks, terrestrial networks, aeronautical networks and oceanic networks into a single unified network covering every part of the world and extending into space}.

As enormous data processing request has emerged with newly proposed applications like ship navigation, positioning, remote real-time sensing, cooperative detection, and information fusion, mmWave technologies are naturally utilized in ground-to-ground, air-to-air, and air-to-ground, air-to-sea links \cite{S1}. However, unlike existing heterogeneous networks, in each layer of SAGS there exists extensive dynamics, posing great difficulty to network planning. Extensive investigations are called for. Hong et al. \cite{C1} noticed the effect of UAV on mmWave channel characteristics in new application scenarios. Di et al. \cite{C2} designed an integrated network architecture to enable network access at both satellite and terrestrial communications. Moreover, the authors of \cite{9171868,9086520} optimized the performance in UAV-aided systems. {\color{black}The recent research \cite{SAGE-2021} laid out the grant vision of the SAGS network for 6G, and discussed in detail new paradigm shifts.} It is clear that mmWave technology will be one of the key enabling technologies for this future generation network.

\section{Mobility Models}\label{S3}

\begin{figure}[!ht]
\vspace*{-2mm}
\begin{center}
\hspace*{-5mm}
\includegraphics[width=\columnwidth]{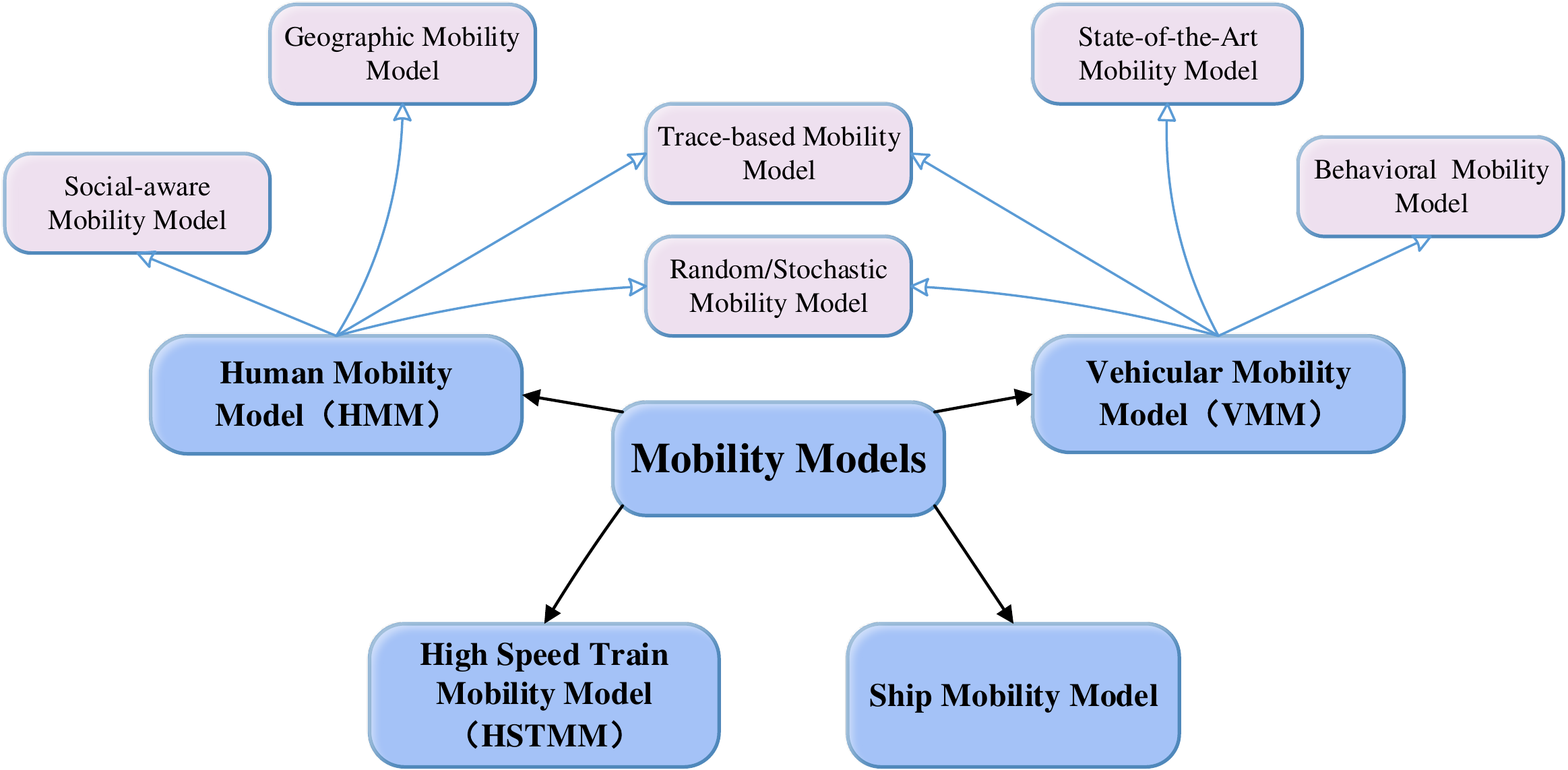}
\end{center}
\vspace*{-2mm}
\caption{\revise{Categories of mobility models relevant to mmWave communications.}}
\label{fig:MMType1} 
\vspace*{-7mm}
\end{figure}

Mobility models as efficient tools to characterize mobile patterns have drawn considerable attention in communication systems \cite{mobility-2016a,mobility-2019a}. Extracted from large-scale data, {\color{black}these models allow researchers to predict the influence of mobility factors: speed, direction, congestion, social interaction, place preference, etc., on network performance \cite{mobility-2016,mobility-2019,mobility-2020}.} Several representative mobility models relevant to mmWave communications are reviewed in this section, \revise{ which include human mobility model (HMM), vehicular mobility model (VMM), high speed train mobility model (HSTMM), and ship mobility model, which are illustrated by Fig. \ref{fig:MMType1}.}
\revise{As an essential part, their applications and new trends are summarized as well.}

\subsection{Mobility Models}\label{S3.1}

\subsubsection{Human Mobility Model (HMM)}\label{S3.1.1}

As mobile communication technologies connect both the physical world and human social life, researchers are increasingly exploiting human mobility properties as a fundamental tool in solving a variety of critical problems ranging from people's behavior observation, social relationships analysis, epidemic spread tracking, etc. Various HMMs \cite{HMMr2019} can be categorized into four types, random models, social-aware models, geographic-based models, and trace-based models.

\emph{a) Random Mobility Model (RMM):}
RMMs capture the random movement patterns of human, and have been widely used in the evaluation and design of mobile networks, including mmWave networks. These models include random waypoint (RW) \cite{RW1996} and its two variants, random walk (RL) \cite{RL2004} and random direction (RD) \cite{RD2017}, as well as Levy walk model (LW) \cite{LW2011}. Among them, RW models initially work as a reference in evaluating mobile ad hoc network (MANET) routing protocols and applications, and they are subsequently used to describe people's mobility with constraints of maximum velocity and pause time \cite{RW2020, RW2019}. Owing to their simplicity, RMMs are frequently used in simulating human mobility but they usually suffer from the drawbacks of speed decay and failing to describe steady status in the simulation. 

To eliminate or mitigate these unrealistic features, RMMs are integrated with temporal dependency and spatial dependency to better {\color{black}realize} randomness and unpredictability. For example, probabilistic random walk (PRW) model \cite{PRW2002} and semi-Markov smooth (SMS) mobility model \cite{sms2006} have been used for years. However, applying RMMs in mmWave mobile communications faces some inherent difficulties because people's movement and communication action are tightly correlated with the environment and social relationship, which cannot be depicted accurately by pure random models. 

\emph{b) Social-aware Mobility Model (SMM):}
Statistic results suggest strong correlations between mobile communication and social network (SN) \cite{social2014,social2018}. SMMs are based on the topological measures of proximity and social interactions for mobile users in SN that {\color{black}reflect} features in both space and time dimensions. In the time domain, human moves in a social context-sensitive manner and may pause for certain social interactions, so that movement duration can be divided into contact time and inter-contact time, respectively, which measures the human encounter frequency and the time interval between encounters \cite{social2014,social2018}. Similarly, in the space domain, humans always show certain habits and preferences for places during their social interactions or daily life \cite{social2014,social2018}. 

\begin{figure}[!tp]
\begin{center}
\includegraphics[width=\columnwidth]{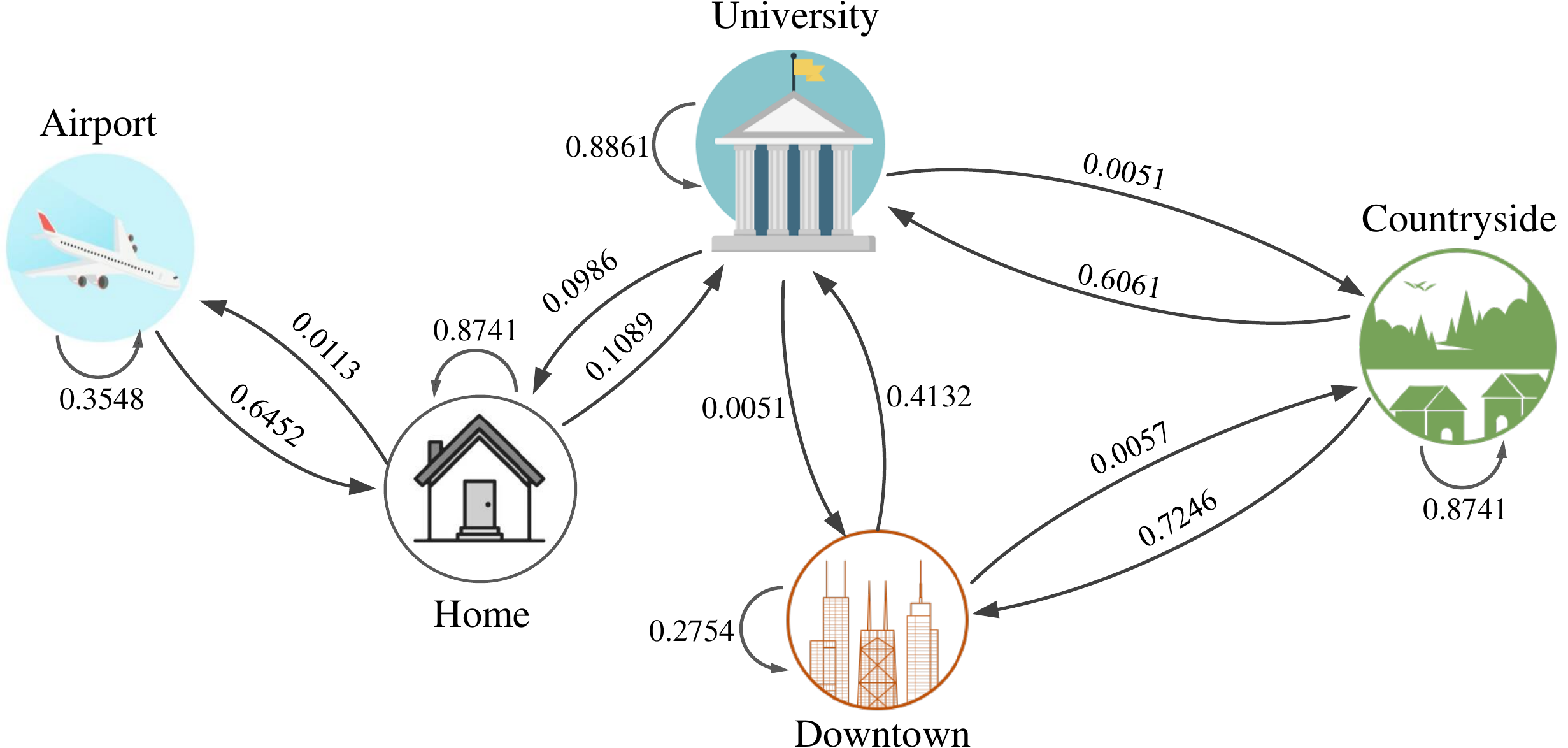}
\end{center}
\vspace*{-2mm}
\caption{A Markov chain based SMM with the representative locations and transition probabilities \cite{Kim}.}
\label{fig:Space} 
\vspace*{-5mm}
\end{figure}

By means of clustering from real data, Fig.~\ref{fig:Space} depicts a Markov chain based SMM, where the representative locations and transition probabilities indicate the location correlation of human traces \cite{Kim}. Essentially, SMMs capture the social nature of humans and are suitable for real-life applications. In particular, as humans are social creatures, they tend to gather in groups and address problems collaboratively. 
Therefore, the community-based mobility models (CMMs) explore deep relationships among people in social communities \cite{social2014,CMM2014}. Since each member in CMM is largely affected by the other users that belong to the group, the co-location information and the relative influences between users are the key parameters to characterize mobility. For example, the encounter frequencies, the human popularity, and online social browse preference make it possible to predict the formation of new social ties \cite{social2014,Social}. SMMs offer higher precision in choosing context to be offloaded and cached, and they are frequently used to enhance the energy efficiency as well as {\color{black}the overall performance} of mmWave networks \cite{social2014,social2018,Social2019a}.

\emph{c) Geographic Restriction Mobility Model (GRMM):}
GRMMs describe humans' movement in bound-aware areas, e.g., campus, shopping center, subway station, etc.  In these areas, movement action may be directed by pathways or be obstructed by location congestion, which indicates that mobility patterns are subject to geographic restriction. There are three categories of GRMMs that can be selected to characterize the movement of mmWave users in real life. 

Path-driven models characterize users' mobility patterns when they are moving along a predefined route or city map \cite{GMM2019}. Event-driven models play an essential role in predicting/characterizing human movement in environmental or local events \cite{GMM2007}. They have wide applications for mmWave communications under accident or disaster situations, where the different roles of people may inspire converse movement. Based on the captured feature among mobility users, properties driven models (PMs) are formed \cite{PM2010}. There are two unique properties in human mobility. Asymptotically, users frequently return to certain locations, such as {\color{black}offices to workers and classrooms to students.} These places are called hot spots. The other one is that the appearance of obstacle nodes interrupts users' predefined route, leading to movement change. These obstruction spots should be integrated into mobility models, while their effect on radio propagation should also be considered. Additionally, physical quantities among mobile users such as distance, spatial cosine similarity, co-location rates are utilized in the foundation of PMs, working together as an optional strategy for mmWave users' mobility description and prediction. 

\emph{d) Trace-based Mobility Model (TBM):}
Detailed analysis shows that the basic statistical properties, such as visiting frequency and popular places, recorded in human mobility traces {\color{black}matter} very much in building accurate human mobility models. This significant finding is documented in \cite{6}. Recent exploration by multiple disciplines have concentrated on real-world traces collection systems, including global positioning system (GPS) \cite{LW2011}, cellular networks \cite{mobility-2019}, and WLAN \cite{HMMr2019} as well as the data processing field that related to machine learning and data mining techniques \cite{mobility-2016}. Due to this progress, TBMs can accurately represent the mobility patterns of mobile users in mmWave scenarios.

\subsubsection{Vehicular Mobility Model (VMM)}\label{S3.1.2}

V2X communications have great potential to enable future intelligent applications, such as smart cities and intelligent transport systems, and exploiting vehicle mobility is of great importance in designing efficient V2X protocols and applications \cite{mobility-2019a,mobility-2020,SOA,VTM2016}. By now, researchers have understood {\color{black}the main features} in various V2X scenarios and have built novel VMMs \cite{mobility-2016a,VMM2,VSN}. Based on the characteristics of models and the priorities of different applications, VMMs can be categorized into the following cases.

\emph{a) Stochastic Model (SM):}
In SMs, vehicles move in a random manner. Although owing to limited interactions between vehicles, SMs, such as Reference Point Group Mobility Model (RPGM), Freeway Model, and  Manhattan Model, have limitations in accurately modeling the complicated vehicular ad hoc network (VANET) applications. However, these models are capable of capturing the stochastic nature of traffic arrivals as well as the complicated movements of vehicles in an ITS \cite{VSM2015}. The stochastic vehicle mobility model of \cite{VMM2} considered {\color{black}the direction and velocity} of the user mobility and was capable of adapting to the traffic condition and type of the street. The work \cite{Lei3} emulated the network throughput with the Manhattan model, and the study \cite{Lei2} leveraged a stochastic VMM in the dynamic optimization of D2D communications. The authors of \cite{VSM2017} considered mmWave networks for highway vehicular communications, where heavy vehicles, like buses and lorries in slow lanes, obstruct LoS paths of vehicles in fast lanes, causing blockages.

\begin{figure*}[!t]
\vspace*{-2mm}
\begin{center}
\hspace*{-5mm}\includegraphics[width=2\columnwidth]{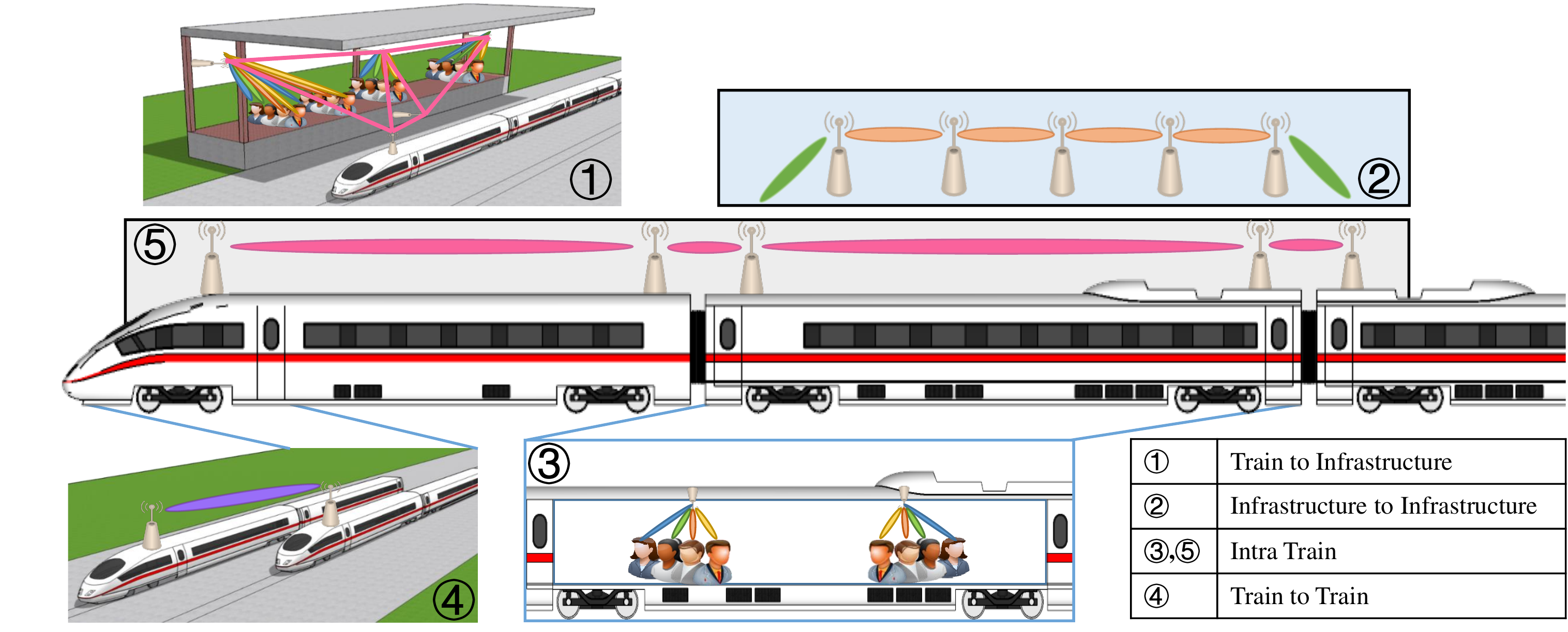}
\end{center}
\vspace*{-3mm}
\caption{HST communication systems: T2I, I2I, T2T, and intra train.}
\label{fig:HST_Full} 
\vspace*{-4mm}
\end{figure*}

\emph{b) Behavioral Model:}
There are two categories of behavioral models. The first category focuses on human behaviors in vehicular scenarios in that they are participants of transportation, playing roles of drivers, pedestrians, or passengers \cite{BM2009}. {\color{black}These models also help to investigate how the human follows traffic advices under emergent situations such as traffic jam and accident \cite{BM2006}.}

{\color{black}The other category believes that the movement of each vehicle is determined by social interactions so that social networks should be exploited in VMMs \cite{social2014}.} Gao et al. \cite{BM2017} investigated the impact of human selfishness on {\color{black}D2D communications underlaying mobile networks.}

\emph{c) Trace-based Model:}
Massive researches have been conducted on designing and collecting vehicle mobility traces, from which important VMMs' parameters can be extracted and the capability of these VMMs can be evaluated \cite{mobility-2016a,VTM2013,VTM2014,VTM2014a}. Based on large-scale real-life vehicular trace data, the study \cite{VTM2013} revealed the exponential and power-law distribution of contact duration in VANETs, which is very different from the case of human mobility.

The work \cite{VTM2014} modeled the macroscopic-level vehicular mobility as a Markov jump process, and used two large-scale urban city vehicular motion traces to validate the proposed vehicular mobility model. Three important metrics related to vehicular mobility and system performance were obtained, which are vehicular area distribution, average sojourn time and average mobility length, and two applications demonstrated the effectiveness of the proposed model in analyzing system-level performance for vehicular networks. By utilizing two large-scale urban city vehicular traces, Li et al. \cite{VTM2014a} proposed an effective vehicular mobility model to analyze the predictability limits of large-scale urban vehicular networks. The findings of \cite{VTM2014a} reveal that there is strong regularity in the daily vehicular mobility, which can be exploited in designing vehicular networks. The connectivity of moving vehicles is one of the key metrics in VANETs that critically influence the performance of data transmission. Hou et al. \cite{mobility-2016a} modeled and investigated the impact of mobility on the connectivity of vehicular networks using a large-scale real-world urban mobility trace. Important findings in this study provide helpful guidelines in the design and analysis of VANETs.

In summary, {\color{black}research on vehicular trace-based mobility models has advanced significantly.} Many of these models and findings can be adopted to mmWave communications in the V2V scenarios.

\emph{d) State-of-the-art Vehicular Mobility Model:}
Since traditional models may not meet {\color{black} well} the different fine-grained requirements in mmWave applications, a promising direction is to construct self-adaptive VMMs. To date, the latest big data analysis \cite{mobility-2019,VTMbd2018} as well as deep learning based techniques \cite{VTMdl2019} have been widely adopted in the VMM research, hence equipping VMMs with big-data driven self-learning capability and enabling a wide range of emerging V2X applications. For example, the work \cite{SOA} proposed the edge-assisted vehicle mobility prediction model (EVM), \revise{which not only adopts a hybrid neural network architecture to process massive mobility data but also allows each vehicle to fine-tune its customized mobility prediction model in a transfer learning manner, thus significantly outperforming traditional models.}

\revise{Moreover, VMMs based on traffic simulators, such as CORSIM \cite{CORSIM1997}, LIMoSim \cite{LIMoSim2017}, PARAMICS \cite{PARAMICS2009}, SUMO \cite{SUMO2011}, TRANSIM \cite{TRANSIM1995}, VISSIM \cite{VISSIM2010} and etc., are growing popular in mobility analysis for mmWave V2X networks. Such models can describe the realistic mobility patterns of different entities in detail (waiting at intersections, turning, crossing, etc.), and therefore they are superior to traditional models in terms of accuracy \cite{NS-3}.}

\begin{table*}[bp!]
{\color{black}
\vspace*{-3mm}
\caption{{\color{black}Applications of Mobility Models in mmWave Mobile Communications, Divided by Model Category}}
\label{table:MMApplication} 
\vspace*{-3mm}
\begin{center}
\resizebox{\textwidth}{!}{
\begin{tabular}{|l|l|l|l|l|l|l|} 
\hline
Ref. & Year & Mobility Model & Frequency & Scenario &Speed & Application \\ \hline\hline
\cite{MMTable5} & 2018 & RW 
 & \begin{tabular}[c]{@{}l@{}}Sub-6 GHz\\ 60 GHz
 \end{tabular} 
 & A two-tier HetNet & [1, 3] m/s & Performance enhancement
\\ \hline
\cite{MMTable4} &2018 &RW &- &- &\begin{tabular}[c]{@{}l@{}}
Walking: 0.83 - 1.388 m/s\\
Running: 4.44 - 6.66 m/s\\
Biking: 4.3 - 11.11 m/s\\
Car: 17.88 - 31.3 m/s. \end{tabular} &Performance degradation analysis 
\\ \hline
\cite{RW2019} &2019 & RW &28 GHz & Peer-to-peer (P2P) networks & uniformly chosen in {\color{black}pre-defined} interval &Coverage analysis
\\ \hline
\cite{RW2020} & 2020 & Orientation-based RW & - 
 & \begin{tabular}[c]{@{}l@{}}Indoor light-fidelity\\
 mmWave cellular networks \end{tabular}
 & 1 m/s, 1.4 m/s, 2 m/s & \begin{tabular}[c]{@{}l@{}}Framework construction for \\performance analysis\end{tabular}
\\ \hline
\cite{MMTableD2D} &2021  &RW &60 GHz &mmWave D2D networks
&\begin{tabular}[c]{@{}l@{}}Pedestrians: [2, 4] m/s\\
Vehicles: [5, 15] m/s \end{tabular}
& Link allocation improvement
\\ \hline\hline
\cite{MMTable7} &2018  &RPGM/RD &73 GHz &A festival or a concert 
& \begin{tabular}[c]{@{}l@{}}UE: 1.4 m/s\\
Drone-cell: 8.3 m/s \end{tabular}
& Network coverage improvement
\\ \hline
\cite{MMTable6}
 & 2019 & RPGM & 60 GHz & mmWave mobile scenarios & 3 km/h & \begin{tabular}[c]{@{}l@{}}mmWave cell-association\\ algorithms comparison\end{tabular}
\\ \hline
\cite{MMTable11} & 2020 & LIMoSim/SUMO/others
 & \begin{tabular}[c]{@{}l@{}}2.1 GHz \\ 5.9 GHz \\28 GHz
 \end{tabular} 
 & {\color{black}Hybrid vehicular networks} & - & \begin{tabular}[c]{@{}l@{}}Mobility simulation framework\\ construction\end{tabular}
\\ \hline\hline
\cite{MMTableN} & 2016 & Self-defining & 60\,GHz & Street canyon 
 & \begin{tabular}[c]{@{}l@{}}Pedestrians: 3-5 km/h \\ Vehicles: 30-120 km/h \end{tabular}
 & \begin{tabular}[c]{@{}l@{}}Evaluating mobility impact on\\ system performance \end{tabular}
\\ \hline
\end{tabular}
}
\end{center}
}
\vspace*{-2mm}
\end{table*}

\subsubsection{High Speed Train Mobility Model (HSTMM)}\label{S3.1.3}

HST as a sustainable ground transportation method {\color{black}has} been developed in many countries. The rapid growth of HST services demands better and more reliable wireless communication systems for the train control
data transmission as well as passenger Internet access and broadband services. The work \cite{HST_KeGuan} investigated the challenges in developing such HST wireless communications. Fig.~\ref{fig:HST_Full} depicts the HST communication system comprising train to infrastructure (T2I), infrastructure to infrastructure (I2I), intra train, and train to train (T2T) links. Defined by the International Union of Railways (UIC) E-Train Project, the T2I wireless systems provide services for train control and monitoring while the intra train wireless systems  offer Internet connections to passenger smart devices \cite{LeiLei}. To optimize mmWave network design in HSTs, HSTMMs are constructed based on mobility features and propagation characteristics associated with HST wireless communication systems. 

Firstly, HSTs always move along pre-constructed tracks, and the location and motion direction of a HST are predictable. Lei et al. \cite{LeiLei} formulated the mobility model of the HST system as a semi-Markov process. {\color{black}The work \cite{HSTMM2019} addressed the challenging task of frequent handover for each communicating user, due to the high mobility of the train.}

Secondly, humans are the main customers for the service of railway systems, and their mobility patterns can be analyzed from two perspectives. At the macroscopic level, thanks to the increasing availability of big data, the authors of \cite{Singapore,Singapore1,BD,46} were able to analyze the client flows with a deeper understanding. For example, Hasan et al. \cite{46} proposed an urban human mobility model for visiting location prediction by observing the smart subway card transactions, while Soh et al. \cite{Singapore1} constructed a complex weighted network for Singapore clients by noticing the traffic flows on hub nodes. From a microscopic perspective, human trails in railway systems may be driven by both observed factors (e.g., schedules, the station facilities) and hidden factors (e.g., social interaction, emergent issues). Modeling these complicated factors with machine learning can be further used to quantify human travel on HSTs.

Thirdly, T2T links illustrated in Fig.~\ref{fig:HST_Full} form crucial safety measures for train control to avoid collision accidents for HSTs running at high speed. There have been some preliminary channel models designed for this purpose \cite{AIBO}. However, further investigations on mobility control are still required.

\begin{table*}[!bp]
\vspace*{-3mm}
\caption{{\color{black}MmWave Mobility Channel Measurement Campaigns, Divided by the Target Network}}
\label{table:Measurement} 
\vspace*{-4mm}
\begin{center}
\resizebox{\textwidth}{!}{
\begin{tabular}{|l|l|l|l|l|l|l|l|l|} 
\hline
Ref. & Year & Target & Frequency & Bandwidth &  Antenna & Mobility Pattern, Speed & Environment & Channel Statistics   \\ \hline\hline
\cite{HRS1} & 2003 & V2V & 60\,GHz & - & SISO, at the roof of vehicle & Multi-mobility, - & \begin{tabular}[c]{@{}l@{}} Highway \\ Regular city road \end{tabular} & PL \\ \hline
\cite{HRS3} & 2017 & V2V & \begin{tabular}[c]{@{}l@{}} 38\,GHz \\ 60\,GHz \end{tabular} & 500\,MHz & SISO, at the bumper & \begin{tabular}[c]{@{}l@{}} Dual-mobility, \\ranging in [40,~70]\,km/h\end{tabular} & Campus & PDP, SF, SSF, DS \\ \hline
\cite{HRS2} & 2018 & V2V & 73\,GHz & 409\,MHz & SISO, at the roof of vehicle & Dual-mobility, 60\,km/h & Urban & PL, SSF \\ \hline
\cite{HRS4} & 2018 & V2V  & 60\,GHz & 510\,MHz & SISO, outside the vehicle  & \begin{tabular}[c]{@{}l@{}} Mono-mobility, \\ ranging in [0,~30]\, km/h \end{tabular} & Urban street & PDP, SF, Doppler \\ \hline
\revise{\cite{60GHzVehicular}} & 2019 & V2V & 60\,GHz & 510\,MHz & \begin{tabular}[c]{@{}l@{}} SISO, \\ TX: aligned towards the RX \\RX: on the left rear car window \end{tabular} & Mono-mobility, 30\,km/h &Urban street &Doppler, CIR \\ \hline
\cite{HSTMeasurement2} & 2019 & V2V & 41\,GHz & 1.25\,GHz & SISO, at the top of vehicle & \begin{tabular}[c]{@{}l@{}} Dual-mobility, \\ {\color{black}exceeding} 170\,km/h \end{tabular} & Suburban street & \begin{tabular}[c]{@{}l@{}} PDP, FD, DS, AF, \\ LCR, AFD \end{tabular} \\ \hline
\cite{MIMOChannel2020} & 2020 & V2V & \begin{tabular}[c]{@{}l@{}} 28\,GHz\\38\,GHz \\ 39\,GHz \end{tabular} & - & MIMO, - & \begin{tabular}[c]{@{}l@{}} Dual-mobility, \\ ranging in [0,~60]\, km/h \end{tabular} & Straight road & PL, PDP \\ \hline\hline
\cite{HST51} & 2018 & HST & 31.625\,GHz & 250\,MHz & \begin{tabular}[c]{@{}l@{}} MIMO, \\TX: along the right side \\ (wall) of the tunnel \\ RX: on the middle of front \\ window in the cab \end{tabular} 
& Dual-mobility, 400\,km/h & Tunnel & PDP, CIR \\ \hline
\cite{Mie} & 2018 & HST & 93.2\,GHz & 2\,GHz & \begin{tabular}[c]{@{}l@{}} SISO, \\ TX: fixed \\ RX: moving along the track \end{tabular} & Mono-mobility, 500\,km/h  & RMa & \begin{tabular}[c]{@{}l@{}}PL, \\ Amplitude statistics \end{tabular} \\ \hline
\cite{HST56} & 2019 & HST & 28\,GHz & - & SISO on the bed of truck & Dual-mobility, 6.3\,km/h & Rural & PDP, DS, AS, K-factor \\ \hline
\cite{HST1} & 2020 & HST & 28 GHz & 500\,MHz & \begin{tabular}[c]{@{}l@{}} SISO, \\ TX: next to the test track \\RX: on the rooftop of the \\ train carriage \end{tabular} & Dual-mobility, 170\,km/h & \begin{tabular}[c]{@{}l@{}} Tunnel \\ Viaduct \end{tabular} & PL, DS, Doppler \\ \hline\hline
\cite{UAVm2020} & 2020 & UAV & 60\,GHz & 2\,GHz & MIMO,TX/RX on UAVs & Dual-mobility, - & A2A, hover, LoS & PL \\ \hline
\multicolumn{9}{|l|}{PL: path-loss, PDP: power delay profile, SF: scatter function, SSF: small scale fading, DS: delay spread, Doppler: Doppler spread, AF: autocovariance function,} \\
\multicolumn{9}{|l|}{FD: fading depth, LCR: level crossing rate, AFD: average fading duration, AS: angle spread, A2A: air-to-air} \\ \hline
\end{tabular}
}
\end{center}
\vspace*{-2mm}
\end{table*}

\subsubsection{Ship Mobility Model}\label{S3.1.4}

Recognition and understanding of ship mobility patterns {\color{black}have} great significance for intelligent maritime applications. The mobility pattern of the ships conducting wireless transmissions is one of the three key factors influencing wireless communication performance in the ocean environment \cite{SMM2007}. To complete the global SAGS network, it is crucial to develop ship mobility models. The authors of \cite{SMM2018} studied the mobility pattern of ships based on the mobility traces of more than 4000 fishing and freight vessels. {\color{black}The results of \cite{SMM2018} provided  useful guidelines on the design of data routing protocols for OANET.} The work \cite{SMM2019} proposed a long-term fine-grained trajectory prediction algorithm for ocean ships, called L-VTP, which takes into account trajectories' sparsity of ocean ships, the different mobility patterns of the same ship during the day and the night. Extensive experiments were conducted based on two years of real-world trajectory data for more than two thousand ships.

{\color{black}
\subsection{Applications and New Trends}\label{S3.2}

Various mobility models reviewed in the previous subsection are all relevant to mmWave mobile communication scenarios. This is because although many of these mobility models predate mmWave mobile communications, mobility models are typically application scenario specific and they are not tied to particular carrier frequency used. Therefore, they have been used in mmWave research. Several representative applications are summarized in Table~\ref{table:MMApplication}, with the relevant references and key system characteristics of frequency, scenario, and mobility speed highlighted. Two findings can be drawn from Table~\ref{table:MMApplication}:

\emph{1)} Although some researchers are investigating the integration of learning-based mechanisms in mobility modeling, simple stochastic/random models are still among the most widely used ones in mmWave research, where behaviors, traces, social relationships, and other specific features of objects are not fully considered.

\emph{2)} Similar to the examples in \cite{MMTable7, MMTable11}, hybrid applications of mobility models have drawn {\color{black}increasing} attention in the upcoming era since a single movement pattern/mobility model can no longer describe the complex mobility mode in HetNets.
}

\section{Key Challenges and Existing Solutions}\label{S4}

The mmWave band covering 30-300 GHz is regarded as a solution to enable Gbps transmission and support emerging mobile applications. However, there are several major technical challenges for mmWave mobile communications in the 5G era and beyond, including channel measurements, channel estimation, anti-blockage method, and so on. 
To understand these challenges deeply, the current problems and existing solutions are discussed concretely in the following.

\subsection{Channel Measurements and Modeling}\label{S4.1}

\subsubsection{Channel Measurements}

Extensive channel measurements help to understand the physical characteristics of {\color{black}mmWave bands}, which are also essential for channel modeling and system design. However, many measurement campaigns were conducted in quasi-static scenarios \cite{BadMeasurement1,BadMeasurement2,InVehicle, ChannelSounder,BadMeasurement3}, and consequently the channel data collected failed to characterize mobile mmWave channels. For example, Blumenstein et al. \cite{InVehicle} measured the static channel impulse response (CIR) at 55-65\,GHz band in {\color{black}the intra vehicle environment}, where the receiving node (RX) and the transmitting node (TX) were fixed inside the vehicle. Similarly, in the measurement experiments of \cite{BadMeasurement2}, TX and RX were relatively static.

With the progress in measurement theory and hardware design, several mobility-aware measurements in {\color{black}mmWave bands} were carried out recently, and the performed measurement campaigns are listed in Table~\ref{table:Measurement}. In this table, the mobility patterns are divided into three types according to the mobility of TXs and RXs as well as scatters, specifically, 1)~\revise{mono-mobility}: one of TXs, RXs, and scatters is mobile; 2)~dual-mobility: two of them are mobile, and 3)~multi-mobility: more than two of them are mobile. It is worth noting that these measurements were mainly conducted in HST and V2X scenarios while MIMO measurements were rare. 
Additionally, much more extensive mobility-aware measurements should be carried out in new application scenarios, including smart agriculture, industrial Internet of Things (IIoT), UAV, SAGS, etc., which provide brand-new services for 5G and beyond (B5G) \cite{6Gm2020}.

\subsubsection{Channel Modeling}

\begin{figure}[!tb]
\begin{center}
\includegraphics[width=\columnwidth]{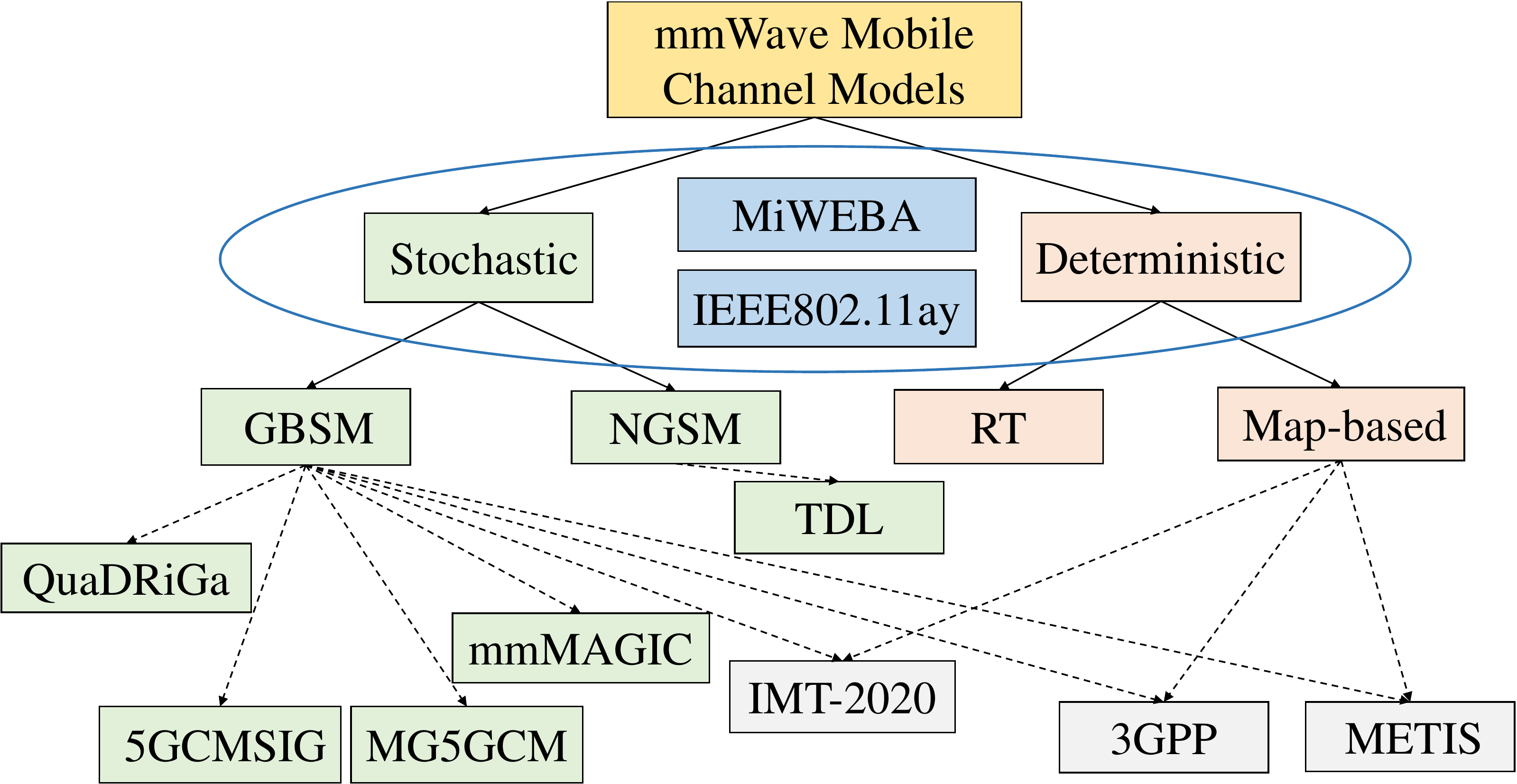}
\end{center}
\vspace*{-3mm}
\caption{Classification of mmWave mobility channel models.}
\label{fig:Classification} 
\vspace*{-6mm}
\end{figure}

Channel models are of significance to the development of mmWave ultra-wideband mobile communications and they are subjected to intensive study. In recent decades, several mobility-support mmWave channel models have been proposed \cite{6Gm2020,2016,Wangchengxiang}, which can be classified roughly into stochastic models and deterministic models. A more detailed classification of various mmWave mobility channel models is provided in Fig.~\ref{fig:Classification}.

Stochastic models describe channel parameters by certain underlying probability distributions, and they are mathematically tractable and can be adapted to various scenarios. As shown in Fig.~\ref{fig:Classification}, they are categorized as two types, geometry-based stochastic model (GBSM) \cite{YinCheng2016} and non-geometrical stochastic model (NGSM) \cite{WCX2018}. GBSMs characterize the propagation environment with mathematical relations among geometric points and clusters. Standard models like QuaDRiGa, 5GCMSIG, MG5GCM, and mmMAGIC are GBSMs, which cover statistical characteristics in different mmWave frequency ranges \cite{Wangchengxiang}. However, GBSMs suffer from two shortcomings. First, since the multipath components from measurement data are extracted by clustering methods, {\color{black}there lacks clear understanding of the multipath physical nature.} Second, measurement data themselves have limitations. This is because {\color{black}the bandwidth} used in measurement campaigns is around 1\,GHz rather than up to 8\,GHz bandwidth as planned by ITU, and most measurements were conducted under quasi-static channels, which are quite different from real mmWave applications. NGSMs characterize the channels in a purely {\color{black}statistical manner} without exploiting geometrical information. Typical NGSMs include tapped-delay-line (TDL) channel models, which model the CIR with taps at certain delays \cite{Recent}. They have been widely used in modeling non-wide-sense-stationary-uncorrelated scattering (non-WSSUS) V2V channels due to {\color{black}their} low complexity and acceptable accuracy \cite{LiYan}.

Deterministic models can reflect certain propagation characteristics of mobile channels accurately, e.g., the large Doppler frequency in high mobility scenarios. In the HST case, lots of the existing research works have difficulty in describing the dual mobility characteristics of the mmWave ultra-wideband channel accurately. \revise {Ray-tracing (RT)} simulation as a tool for propagation prediction becomes popular in HST communication systems. Based on the geometrical optics (GO) theory and uniform theory of diffraction (UTD), RT is capable of studying {\color{black}multipath phenomena} caused by reflection, scattering, diffraction and can provide solutions to spatial characteristics collection \cite{Hur1}. Thus, RT offers a promising modeling approach to future wireless communications, owing to its three advantages. 1)~Compared to the measurement-based random channel modeling, RT modeling is less affected by the bandwidth and frequency band, enabling its use in the study of channel characteristics from sub-6 GHz to the THz band. 2)~The output of the RT simulator provides high spatial resolution, satisfying the requirements for high channel resolution in beamforming and beam-tracking. 3)~Due to the limitations of hardware equipment, permission issues, and manpower scheduling, measurement campaigns in time-varying and MIMO channels encounters great difficulties. RT simulation helps to mitigate these difficulties and to offer a scenario-specific solution. Another class of deterministic models are map-based models \cite{Lim2020}, which are obtained using RT methodology in a simplified three-dimensional (3D) scene of a propagation environment, and have a nature of spatial consistency. Nevertheless, the key obstacle in deterministic modeling is that it is computationally intensive and its accuracy highly depends on the modeling scenario. 

From the above discussion, it can be seen that stochastic models generally have low complexity but {\color{black}less accuracy}, while deterministic models have better accuracy but are computationally expensive. Therefore, it is highly desired to derive quasi-deterministic (Q-D) modeling methods by combining both stochastic and deterministic approaches, which enjoys the advantages of both stochastic and deterministic models. 
\revise{Specifically, based on the mmWave CIR representation, this hybrid approach models the Q-D strong rays (D-rays) in a deterministic manner as well as models the relatively weak random rays (R-rays), originating from the static surfaces reflections, and the flashing rays (F-rays), originating from dynamic objects reflections, in a stochastic method. Thus a Q-D model is no longer {\color{black}highly dependent on} the detailed scenario description and achieves higher accuracy than pure statistical models \cite{QD1, QD3}.}
Fig.~\ref{fig:Classification} lists two Q-D channel models: MiWEBA \cite{MiWEBA2014,QD1} and IEEE 802.11ay \cite{Maltsev2016}, which support outdoor channel modeling at 60\,GHz. Besides, GBSM and map-based modelings can be used together in IMT-2020 \cite{IMT2020-2019}, 3GPP \cite{3GPP2017}, and METIS \cite{METIS2015}. 

\subsection{Channel Estimation}\label{S4.2}

With the explosively increasing requirements for data exchange in mobile communications, MIMO technique and hybrid network architecture are necessary for mmWave systems. However, the synchronization among multiple antennas in a complex network makes it challenging to obtain accurate channel estimation (CE) \cite{XiaoMing}. To cope with this problem, Alkhateeb et al. \cite{Estimation3} proposed a mmWave CE method that exhibits superior performance in complicated multipath channel environments, making it applicable to multi-flow multiplexing scenarios. By exploiting {\color{black}the} compressive sensing (CS) technique and hybrid precoding method, Al-Nimrat et al. \cite{Estimation1} proposed a low-complexity CE scheme for mmWave Massive MIMO systems in the dense urban environment. Specifically, by analyzing the sparse nature of multipath components (MPCs), the authors of \cite{Estimation1} designed a transmission model and a precoding scheme with a combination of matched filter (MF)/zero-forcing (ZF)/minimum mean square error (MMSE) to improve the system capacity. Liao et al. \cite{mWadd4} developed a closed-loop {\color{black}(CL)} sparse CE scheme for wideband mmWave full-dimensional massive MIMO systems, which harnesses the channel sparsity in both angle and delay domains. This CE scheme is capable of acquiring the super-resolution estimates of both the uplink and downlink angles of arrival (AoAs)/angles of departure (AoDs) and delays of sparse MPCs as well as the least-squares estimates of the path gains with low training overhead. Compared with the existing state-of-the-art CS-based CE schemes \cite{mWCS2016,mWCS2017,mWCS2017a,mWCS2017b,mWCS2018}, the solution of \cite{mWadd4} offers better CE performance while imposing lower computational complexity. 

In recent years, considerable research efforts have been focused on combing machine learning (ML) with beamspace CE for mmWave MIMO CE. The related works \cite{E4,E5,DL-CE2019,DL-CE2020,DL-CE2020a} have revealed that applying machine learning tools, like Bayesian learning, deep learning, etc., is capable of designing robust and adaptive CE mechanisms suitable for time-varying MIMO channels, which outperform their more conventional signal processing based counterparts. For example, Zhang et al. \cite{DL-CE2020a} proposed a fully convolutional  denoising  approximate message passing (FCDAMP) algorithm for mmWave massive MIMO systems, which attains more accurate CE and higher achievable sum rate, especially under low-SNR conditions. Moon et al. \cite{DL-CE2020} proposed a deep learning-based CE and tracking algorithm for vehicular mmWave communications. Specifically, for CE, the authors applied a deep neural network to learn the mapping function between the received omni-beam patterns and mmWave channel with small overhead.

For fast-changing mobile channels, the channel tracking (CT) becomes necessary, which exploits the temporal correlation and supports real-time updating for channel status information (CSI). Common CT methods include improved beam tracking, {\color{black}data-aided} and geometric relationship based schemes \cite{XiaoMing}. Beam tracking methods \cite{CT111}, including Kalman filtering (KF)-based channel tracking, extended Kalman filtering (EKF)-based beam tracking, and least mean square (LMS)-based beam tracking, have lower computational complexity and are explored to track channel parameters, such as AoA and CSI. Furthermore, the power of deep learning can be harnessed for CT. For example, the work \cite{DL-CE2020} applied the long short-term memory (LSTM) network to track the channel, after the initial CE.

\revise{Table~\ref{table-CE} summarizes the literature review for mmWave CE applications.}
In summary, CE and CT schemes incorporating with the state-of-art techniques are important to signal detection and demodulation procedures in mmWave mobile communications. Additionally, although the application of MIMO leads to synchronization issues, the correlated sparse nature of mmWave massive MIMO channels in both angle and delay domains is worth further investigating. 

\begin{table}[h!]
{\color{black}
\caption{MmWave Channel Estimation Applications}
\label{table-CE} 
\vspace*{-4mm}
\begin{center}
\resizebox{0.5\textwidth}{!}{
\begin{tabular}{|l|l|l|} 
\hline
Ref. & Year & Application \\ \hline
\cite{Estimation3}     & 2014 & Hybrid {\color{black}precoding} aided CE for mmWave cellular \\ \hline
\cite{mWCS2016} & 2016 & Hybrid {\color{black}precoding} aided CE for wideband mmWave \\ \hline
\cite{mWCS2017}  & 2017 & CE for wideband mmWave system \\ \hline
\cite{mWCS2017a}  & 2017 & CE for wideband mmWave MIMO \\ \hline
\cite{mWCS2017b}  & 2017 & Sparse CE for mmWave massive MIMO \\ \hline
\cite{mWCS2018}  & 2018 & CS based CE for wideband mmWave MIMO \\ \hline
\cite{mWadd4}  & 2019 & CL sparse CE for wideband mmWave MIMO \\ \hline
\cite{Estimation1}   & 2019 & Low complexity CE for mmWave massive MIMO \\ \hline
\cite{DL-CE2019} & 2019 & ML based mmWave massive MIMO CE \\ \hline
\cite{DL-CE2020} & 2020 & ML based CE \& CT for mmWave vehicular \\ \hline
\cite{DL-CE2020a} & 2020 & ML based beamspace CE for mmWave massive MIMO \\ \hline
\cite{CT111} & 2020 & Beam tracking/CT for mobile mmWave networks \\ \hline
\end{tabular}
}
\end{center}
}
\vspace*{-5mm}
\end{table}

\subsection{Anti-Blockage}\label{S4.3}

As mmWave signals suffer from high penetration loss, communication networks are vulnerable to \revise{dynamic} blockage, which may cause link interruptions and lead to loss of data. 
\revise{The study \cite{MobileBlockers} revealed that dynamic blockage in the environment may introduce sharp drops (up to 30-40\,dB) to the received signal strength. Therefore, it is crucial to account for the blockage effect in performance evaluation or network planning for mobile mmWave communications. The authors of \cite{MIMOChannel2020} studied the blockage effect of human body and vehicle for mmWave signal, giving the lower bound and upper bound of the attenuation based on the knife-edge diffraction (KED) model and geometrical theory of diffraction (GTD) model. The work \cite{MobileBlockers} modeled human body blockage in a moving mmWave system. As illustrated in the work \cite{ms}, the average blockage duration in a highway scenario can range from 100\,ms to even a few seconds. These studies provide meaningful insights and guidance for future mmWave indoor hotspot and vehicular network applications.}

\begin{figure}[!h]
\vspace*{-4mm}
\begin{center}
\includegraphics[width=\columnwidth]{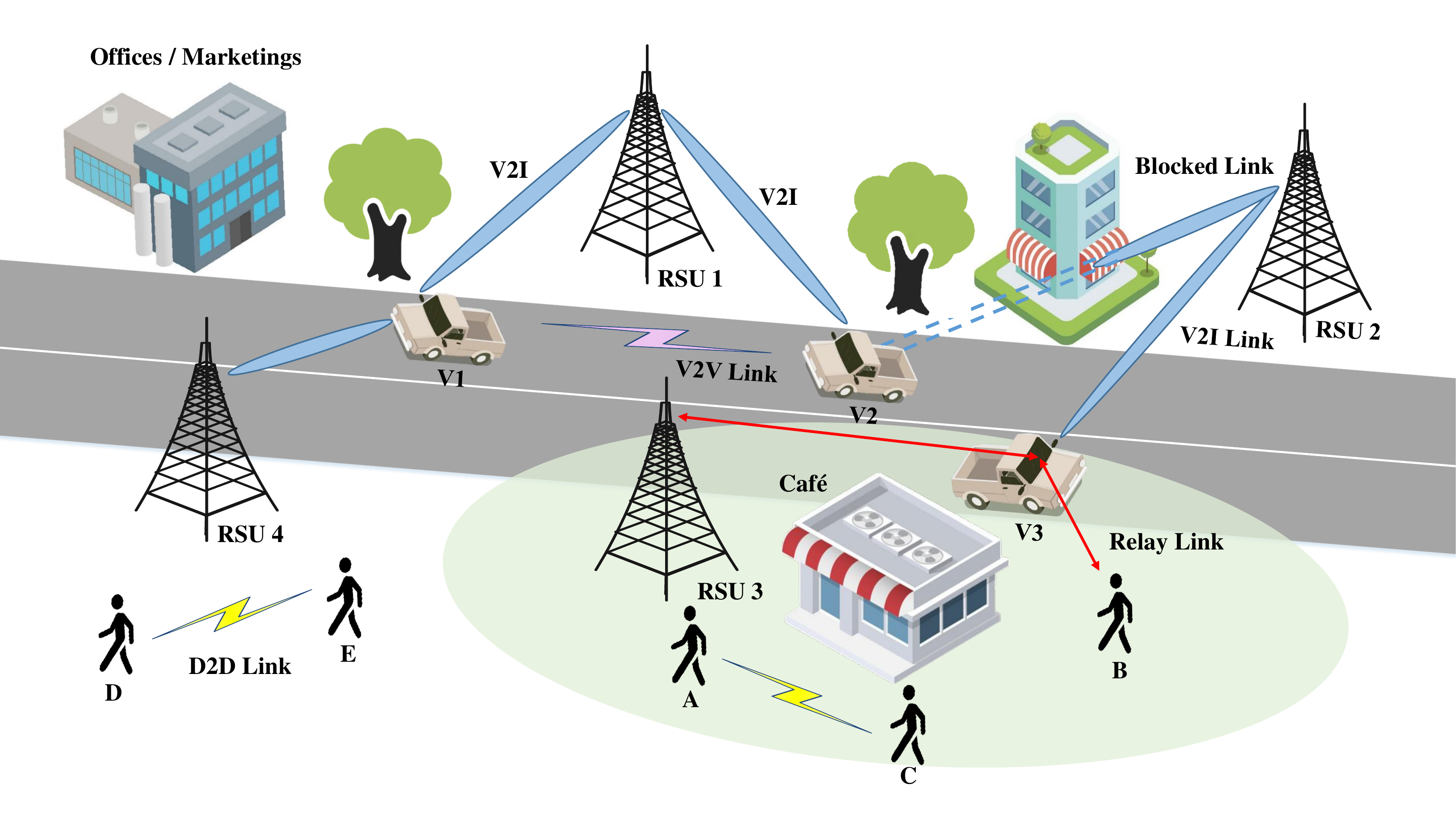}
\end{center}
\vspace*{-6mm}
\caption{Examples of mmWave connection maintenance techniques.}
\label{fig:Antiblockage} 
\vspace*{-1mm}
\end{figure}

\revise{More importantly, different solutions have been proposed to maintain reliable connections in mobile scenarios.} 
Fig.~\ref{fig:Antiblockage} illustrates three main categories of anti-blockage approaches, including multi-connectivity (MC), beamforming (BF), and relay assistance. 

\subsubsection{Multi-Connectivity}

Multi-connection as an available approach for session maintenance in mobile mmWave networks, has been standardized by 3GPP {\color{black}\cite{mc-3GPP}}. MC enables the UE to access multiple BSs simultaneously so that data transmission can be maintained, even one of the connections is interrupted by the blockage. Taking vehicle V2 in Fig.~\ref{fig:Antiblockage} as an example, although its connection to roadside unit 2 \revise{(RSU2)} is blocked, it can continue to communicate with RSU1. Obviously, network capacity as well as outage probability can be improved in a MC-aided mmWave system. However, how to balance the system complexity and the achievable performance remains a problem that requires further exploring. On the analytical level, the work \cite{Upper} offered a closed-form upper bound on the cumulative distribution function \revise{(CDF)} of capacity for the mmWave cellular system supporting MC capabilities, which can be utilized as a benchmark result for the performance evaluation in realistic scenarios. {\color{black}On the technical level, extensive works \cite{mCA_11,Picture1,MC3,MC2021} assessed the indicators related to performance optimization, which covered the MC structure design, strategies selection, deployment density and resource allocation.}

Envisaged programmable MC offers potential for application-level resource scheduling, serving for quality of experience improvement \cite{MC4}. With the help of statistical theory, queueing theory, and powerful simulation tools, MC as a promising technology is capable to assess session-level dynamics of typical mmWave deployments \cite{mCA_11}. 
{\color{black}Table~\ref{table-MC} summarizes the related research on mmWave MC.}
However, current research is mainly conducted in the urban environment, and there are more typical application scenarios that are worth investigating in the future \cite{V2IMC}. Besides, the performance evaluation on scenario-specific upper-layer protocols should be considered in realistic systems \cite{Picture1}.

\begin{table}[th!]
{\color{black}
\caption{{\color{black}Research on MmWave Multi-Connection}}
\label{table-MC} 
\vspace*{-4mm}
\begin{center}
\resizebox{0.5\textwidth}{!}{
\begin{tabular}{|l|l|l|} 
\hline
Ref. & Year & {\color{black}Technique/Contribution} \\ \hline
\cite{MC3}      & 2015 & Performance evaluation of MC in mmWave 5G \\ \hline
\cite{mCA_11}   & 2017 & Dynamic MC ultra-dense urban mmWave deployments \\ \hline
\cite{Upper}    & 2018 & Capacity analysis of mmWave 5G cellular with MC \\ \hline
\cite{Picture1} & 2019 & MC in mmWave 5G cellular urban deployments \\ \hline
{\color{black}\cite{MC2021}} &{\color{black}2021} &{\color{black}\begin{tabular}[c]{@{}l@{}}MC enabled user association and power allocation in\\ mmWave networks\end{tabular}}\\ \hline
\end{tabular}
}
\end{center}
}
\vspace*{-2mm}
\end{table}

\begin{table*}[bp!]\setcounter{table}{9}
\vspace*{-3mm}
{\color{black}
\caption{\revise{Applications of Hybrid Beamforming in mmWave Mobile Communications}}
\label{table:BFTable} 
\vspace*{-4mm}
\begin{center}
\resizebox{\textwidth}{!}{
\begin{tabular}{|l|l|l|l|l|l|l|l|} 
\hline
Ref. & Year & Technique & Frequency & Bandwidth & Scenario & Speed & Application    
\\ \hline
\cite{HBF2018} & 2018 & HBF and CoMP & 28 GHz & 100 MHz & Multi cells & - & Spectral efficiency improvement \\ \hline
\cite{BF5} & 2018 & Adaptive multi-beamforming & 38 GHz & 1 GHz & HST & 100 m/s & Capacity improvement
\\ \hline
\cite{BF2} & 2019 & Joint static/dynamic subarray scheduling & 60 GHz & 2.16 GHz & Mesh backhaul & - & Throughput improvement \\ \hline
\cite{BFWCNC} & 2019 & HBF and task allocation & 28 GHz &50 MHz &- &- & Time delay reduction
\\ \hline
\cite{Gaomeilin} & 2020 & HBF and multi-user MIMO & 32 GHz & 500 MHz & HST & -  & Anti-blockage
\\ \hline
\cite{3DBF} & 2020 & 3D beamforming & 26 GHz & 400 MHz & UAV & 14 m/s & Handover rate reduction \\ \hline
\cite{BFTITS} & 2021 & Adaptive beamforming & 60 GHz & 1.08 GHz & Highway & - & \begin{tabular}[c]{@{}l@{}}Improving efficiency of broadcasting\\ messages\end{tabular} \\ \hline
\end{tabular}
}
\end{center}
}
\vspace*{-2mm}
\end{table*}

\subsubsection{Beamforming}

As an alternative anti-blockage method, BF steers the majority of signals generated by the transmitting antenna array toward an intended angular direction, forming the directional beam to mitigate the interference effect, enhance the transmission robustness, \revise{and achieve superior performance \cite{BF1, BF2016}.} 
BF is growing popular in mmWave mobile scenarios like V2X, HST and UAV. 
In \cite{Feedback1}, a random beamforming scheme suitable for the fast {\color{black}time-varying} situation has been designed for mmWave non-orthogonal multiple access (NOMA) transmission. This approach yields significant performance gains while reducing the amount of feedback to one bit.

\begin{figure}[!htp]
\vspace*{-3mm}
\begin{center}
\includegraphics[width=\columnwidth]{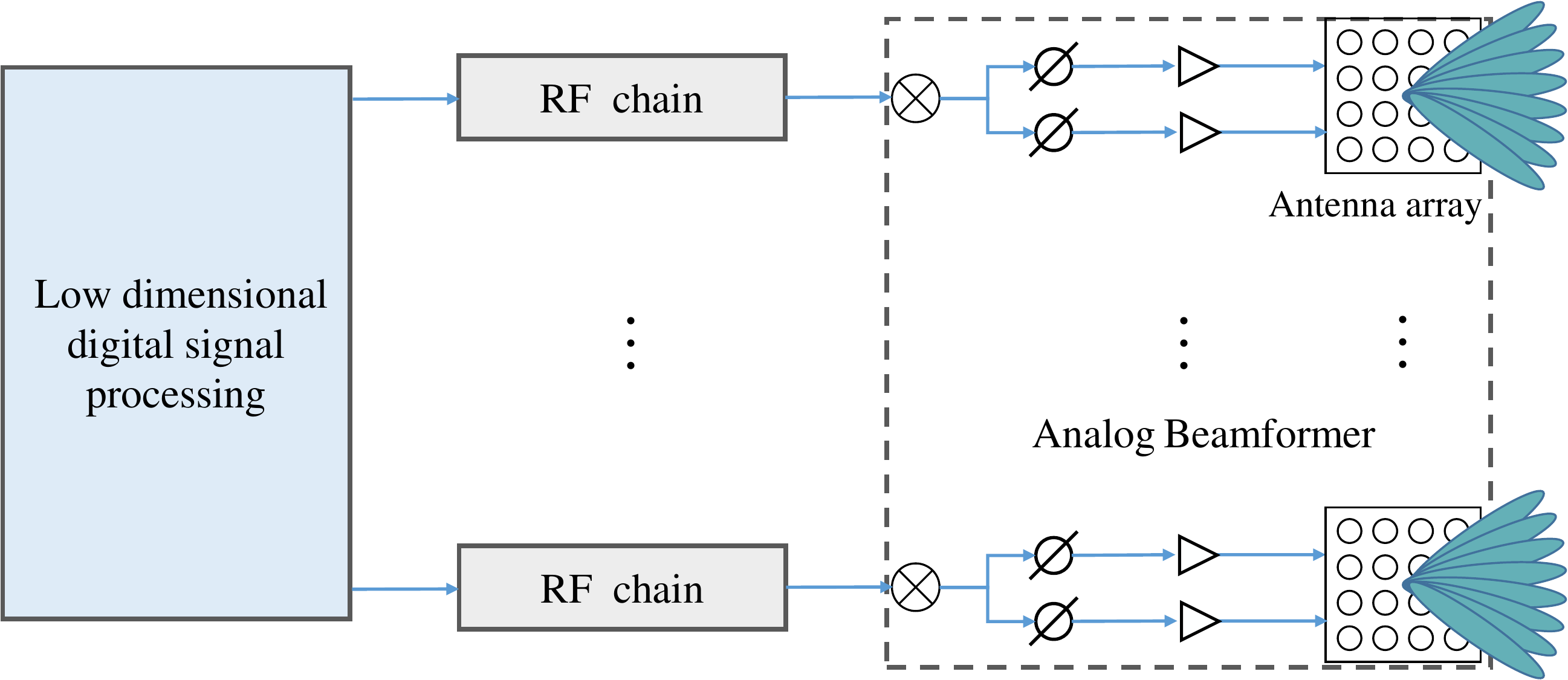}
\end{center}
\vspace*{-3mm}
\caption{Hybrid Beamforming.}
\label{fig:MIMO} 
\vspace*{-3mm}
\end{figure}

In practice, most of mmWave systems rely on large-scale antenna arrays for high beamforming gains, and consequently  powerful full digital BF is impractical as it requires a RF chain for each antenna element. Therefore, hybrid beamforming (HBF) is particularly relevant in mmWave applications, which can combine both the advantages of digital BF in the baseband/digital domain and analog BF in the RF/analog domain. HBF is illustrated in Fig.~\ref{fig:MIMO}, where the analog stage works to generate high beamforming gains from the large antenna array while the digital part can implement digital precoding to support multiple data streams with {\color{black}a} very small number of RF chains. This allows designers to deploy a very large number of antenna elements for the required high beamforming gains, while reducing energy consumption and system complexity.

{\color{black}Many researchers} \cite{BF3,BF2,HBF2019,HBF2019a,HBF2018,HBF2021} have focused on various HBF solutions for the mmWave architecture configuration, signal processing, RF system implementation, etc. More specifically, Roh et al. \cite{BF3} conducted a feasibility study of HBF for mmWave 5G, while Zhai et al. \cite{BF2} studied HBF for mmWave backhaul networks. The work \cite{HBF2019} studied the HBF based on the MMSE design for mmWave systems, and the work \cite{HBF2019a} proposed a hardware-efficient HBF design for mmWave MIMO. Furthermore,  the work \cite{HBF2018} proposed a HBF design for multi-cell multi-user multi-stream mmWave systems by leveraging coordinated multipoint (CoMP), while performance analysis of HBF for multi-user mmWave massive MIMO systems was provided in \cite{HBF2021}.

\begin{table*}[tp!]
\vspace*{-1mm}
\caption{Relay Applications in mmWave Mobile Communications}
\label{table:Relay} 
\vspace*{-4mm}
\begin{center}
\resizebox{\textwidth}{!}{
\begin{tabular}{|l|l|l|l|l|l|l|l|} 
\hline
Ref. & Year  & Relaying Device & Frequency &{\color{black}Bandwidth} & Scenario & Application \\ \hline
\cite{RelayJSAC} & 2009 & Group of devices (DEVs) & 60\,GHz &{\color{black}-}& {\color{black}mmWave WPAN} & Throughput improvement \\ \hline
\cite{RelayJ1}& 2015 & User devices & 60\,GHz &{\color{black}1.7\,GHz} &{\color{black}mmWave WPAN} & Cooperative multicast \\ \hline
\cite{Relay4}    & 2016 & Relay station & 28\,GHz &{\color{black}1\,GHz} &{\color{black}mmWave cellular network} & Energy efficiency improvement \\ \hline
\cite{Relay5}    & 2018 & Vehicle &- &{\color{black}-}& {\color{black}mmWave vehicular network} & Coverage expansion  \\ \hline
\cite{Relay2}    & 2019  & UE & 60\,GHz &{\color{black}-}&{\color{black}mmWave cellular network} & {\color{black}\begin{tabular}[c]{@{}l@{}}Deep learning based relay selection for\\ anti-blockage\end{tabular}}\\ \hline
\cite{Relay6}    & 2019 & UAV & 28\,GHz &{\color{black}1\,GHz}& {\color{black}UAV-enabled mmWave network} & {\color{black}Reliability improvement} \\ \hline
\cite{Gaomeilin} & 2020 & Mobile relays (on top of train) & 32\,GHz &{\color{black}500\,MHz}& {\color{black}mmWave HST network} &{\color{black}HBF design for anti-blockage} \\ \hline
\cite{Relay3}    & 2020 & Vehicle & - &{\color{black}-}& {\color{black}mmWave vehicular network} & Power allocation \\ \hline
\cite{MMTableD2D} & 2021 & UE & 60\,GHz &{\color{black}-}&{\color{black}mmWave D2D communication}& Relay selection by obstacle learning \\ \hline
\end{tabular}
}
\end{center}
\vspace*{-5mm}
\end{table*}

The state-of-art HBF helps to meet the specific requirements of mmWave applications, including throughput, quality of service (QoS), latency, sum rate, etc. For example, the study \cite{Gaomeilin} designed a two-phase algorithm to perform sum-rate maximization. The first phase realizes a feasible optimal beamformer in the blockage-free scenario, and the second phase invokes different strategies to tackle different blockage scenarios. Likewise, the works \cite{BF5,BF2} leveraged joint or adaptive HBF schemes to optimize throughput, power consumption, and antenna gain.
\revise{Table~\ref{table:BFTable} summarizes some key applications of HBF for mmWave mobile networks.}
However, several problems, including hardware limitations, fast time-varying channel, beam alignment and frequent handover, still exist in BF applications, and how to obtain desired performance-complexity trade-off remains a challenging task. 

\subsubsection{Relay}

As another anti-blockage alternative, relay-aided communications help to circumvent obstacles and extend coverage as well as to save transmit power and offer higher data rates than direct links. Fig.~\ref{fig:Antiblockage} shows an example of relaying, where RSU3 attempts to communicate with person B but the direct transmission link is blocked by the Cafe. Relaying through V3 provides an alternative path for communication. Table~\ref{table:Relay} lists various relay applications in mmWave mobile communications, in which relay nodes include BSs, vehicles, \revise{UAVs} and one or a group of UE. 

Mobility behaviors of communication entities may cause frequent handover among relay stations, resulting in complex scheduling, increased energy consumption, potential delay, and even link interruptions. To address these problems, there are strict requirements for relay selection, placement, and scheduling. The design requirements of these issues are often inherently connected. For example, \cite{SAA} optimized the time-slot level throughput by deploying a dynamic relay positioning policy and designing a tractable beamforming approach. Likewise, the work \cite{RelayJ1} achieved power saving and robustness enhancement by proposing a joint solution for relay selection and power allocation under mixed LoS and NLoS conditions.
Also, some researchers address these problems not through hybrid strategies but based on obstacle analysis. For instance, considering that the presence of dynamic obstacles at surroundings causes unpredictable fluctuations to channel quality, the work \cite{MMTableD2D} proposed a relay selection scheme for D2D communications through obstacle learning, in order to assign smart links. Likewise, based on the captured uncertainty introduced by dynamic obstacles, a simple stationary policy is derived in \cite{RelayICC} to guide relay switch {\color{black}decisions}.

In summary, anti-blockage techniques have emerged as key solutions to provide link maintenance, effective coverage, and dynamic capacity in mmWave communication systems. 

\subsection{Capacity Enhancement}\label{S4.4}

MmWave communications should be mobility-adaptive with the aid of various techniques, to achieve efficiency and reliability, and especially to maintain network capacity. Two key aspects of capacity-aware research are discussed below.

\begin{table*}[tp!]
\vspace*{2mm}
{\color{black}
\caption{Scheduling-Based Throughput Optimization for Mobile mmWave Applications}
\label{table-TO} 
\vspace*{-4mm}
\begin{center}
\resizebox{\textwidth}{!}{
\begin{tabular}{|l|l|l|l|l|l|} 
\hline
Ref. & Year & Technique &{\color{black}Frequency} &{\color{black}Bandwidth}& {\color{black}Application Scenario}  \\ \hline
\cite{TH1}   & 2018  & Mobility-aware throughput-efficient service scheduling  &{\color{black}-} &{\color{black}-} & mmWave cells\\ \hline
\cite{D2Dmulticasting} & 2019 & {\color{black}\begin{tabular}[c]{@{}l@{}}D2D communication enabled multicast scheduling for improving\\ throughput and energy efficiency\end{tabular}}&{\color{black}60\,GHz}&{\color{black}2.16\,GHz}& mmWave cells \\ \hline
\cite{schedul2019}     & 2019 & {\color{black}\begin{tabular}[c]{@{}l@{}}Distributed and network-coordinated beam scheduling for sum\\ rate improvement\end{tabular}}&{\color{black}28\,GHz} &{\color{black}-} & mmWave cellular network     \\ \hline
\cite{schedul2020}     & 2020  & {\color{black}\begin{tabular}[c]{@{}l@{}}Joint relaying and spatial sharing multicast scheduling for improv-\\ing reachability, link rate and spatial gain\end{tabular}} &{\color{black}73\,GHz} &{\color{black}1\,GHz} & mmWave cellular network  \\ \hline
\cite{TPopti2020}      & 2021  & Joint time and power allocation for throughput maximization &{\color{black}28\,GHz} &{\color{black}1\,GHz}& Multi-UAV enabled mmWave WPCN  \\ \hline
\end{tabular}
}
\end{center}
}
\vspace*{-6mm}
\end{table*}

\subsubsection{Throughput Optimization}

Mobility imposes serious challenges to mmWave communications. For example, by sharing spectral resources with cellular communication, D2D communication exploits good local channel quality to offer high throughput services. But complex interference induced in mobile D2D links may on the other hand decrease the system capacity. {\color{black}Extensive research has focused on} interference mitigation. By minimizing the mutual interference (MUI) among D2D and cellular users, the joint resource allocation scheme of \cite{2015} maximized the total throughput in the network. Reducing MUI enables concurrent transmission while self-interference (SI) cancellation makes it possible for full-duplex (FD) systems, which incur potential gain in mmWave mobile systems. Users' mobility information is utilized by Yang et al. \cite{LYang} to perform capacity optimization. Specifically, by capturing the distribution regularity of mobility users' popular contents, the authors proposed a low-complexity algorithm for downloading. Similarly, the work \cite{SocialAware} demonstrated that the D2D multicast scheme produces more effective transmission by utilizing both the physical and social properties of mobile users, resulting in throughput maximization and fair allocation of the overall network. In future research, content-sharing intelligent D2D communication will receive more attention and mobility characteristics will play an increasingly important role \cite{mobility-2019a,D2Dopti2014a,TPopti2020}. 
{\color{black} For example, the work \cite{TPopti2020} leveraged multi-UAVs for wireless powered communication network (WPCN), to jointly optimize transmit power and energy transfer time.}

\begin{table*}[bp!]
\vspace*{-3mm}
\caption{\revise{Energy Efficiency Optimization for Mobile mmWave Applications}}
\label{table-EE} 
\vspace*{-4mm}
\begin{center}
\resizebox{\textwidth}{!}{
\begin{tabular}{|l|l|l|l|l|l|} 
\hline
Ref. & Year & Technique &{\color{black}Frequency} &{\color{black}Bandwidth} & {\color{black}Application Scenario} \\ \hline
\cite{JSACUA} & 2017 & Energy efficient user association and power allocation &{\color{black}60\,GHz}&{\color{black}1.2\,GHz}& {\color{black}mmWave
based ultra dense network} \\ \hline
\cite{Green} & 2017 & {\color{black}\begin{tabular}[c]{@{}l@{}}Minimizing energy consumption via concurrent scheduling and\\ power control\end{tabular}} &{\color{black}60\,GHz}&{\color{black}2.16\,GHz}& {\color{black}mmWave backhaul network}\\ \hline
\cite{Precoding} & 2018 & Energy efficient hybrid precoding &{\color{black}60\,GHz}&{\color{black}-}&  mmWave multi-user system \\ \hline
\cite{NiuLi} & 2018 & D2D enabled efficient multicast scheduling &{\color{black}60\,GHz}&{\color{black}-}& {\color{black}mmWave HCN} \\ \hline
\cite{Jingli} & 2019 & Contention graph based energy-efficient FD concurrent scheduling &{\color{black}60\,GHz}&{\color{black}2.1\,GHz}& {\color{black}mmWave backhaul network} \\ \hline
\cite{Power} & 2019 & Discontinuous reception for EE and link reliability &{\color{black}\begin{tabular}[c]{@{}l@{}}28\,GHz\\ 140\,GHz\end{tabular}}&{\color{black}400\,MHz}& mmWave and THz systems \\ \hline
\cite{EE-D2D2020} & 2020 & D2D-enabled multicast scheduling to minimize energy consumption &{\color{black}60\,GHz}&{\color{black}1\,GHz}& {\color{black}mmWave cellular network}  \\ \hline
\cite{NOMAUAV} & 2021 & Optimizing UAV placement, hybrid precoding and power allocation &{\color{black}-}&{\color{black}-} & mmWave NOMA-UAV network \\ \hline
\end{tabular}
}
\end{center}
\vspace*{-2mm}
\end{table*}

Mobility also causes uncertainties for scheduling in multi-hop communications, especially in multi-cell scenarios. Liu et al. \cite{TH1} proposed a throughput-efficient service scheduling scheme, but it is unsuitable for delay-sensitive services. Some researches have addressed the scheduling problem from the network viewpoint. The work \cite{D2Dmulticasting} proposed an efficient multicast scheduling for D2D communications in mmWave small cells. The work \cite{schedul2019} designed a distributed coordinated beam scheduling to mitigate inter-cell interference, which does not require any information exchange between the user and the BS. The authors of \cite{schedul2020} designed highly efficient multicast scheduling for mmWave networks by jointly exploiting the relaying and spatial sharing gains.
Table~\ref{table-TO} summarizes some key schemes of {\color{black}scheduling-based} throughput optimization for mobile mmWave applications.

\subsubsection{Energy Efficiency}

As the transmission demand for data streaming increases tremendously, power consumption has become a critical issue in mobile communication networks. Choosing a short transmission path between TX and RX is a typical way to enhance energy efficiency (EE) in mmWave networks. However, since mobility introduces many varying factors, e.g., flow delay, dynamic topology, frequent switching, etc., into the system, there is no guarantee that choosing a short path will always lead to energy saving. What we need are practical mechanisms for EE performance optimization. Up to now, such mechanisms have been carefully designed,  
most of which are realized by solving the associated EE optimization problems.
\revise{For example, the authors of \cite{JSACUA} focused on power allocation and user association to achieve EE improvement.}
The work \cite{Precoding} proposed the use of hybrid precoding to maximize EE in mmWave multi-user systems. 
Similarly, the work \cite{NOMAUAV} developed an EE enhancement design for {\color{black}the} mmWave NOMA-UAV network by optimizing the UAV placement, hybrid precoding and power allocation.

Notice that for complicated joint design problems with strict constraints, mathematical tools, including graph theory, game theory, convex optimization, deep learning, queuing theory, etc., are widely used to obtain the solutions. For example, leveraging subchannel grouping, the work \cite{Ge} proposed a closed-form EE solution for the MIMO orthogonal-frequency-division multiplexing (MIMO-OFDM) mobile system with QoS constraint. The work \cite{NiuLi} designed an efficient multicast scheduling scheme for mmWave small cells, referred to as CONMD2D, which allows concurrent transmissions. Compared to standard time division multiple access (TDMA), the CONMD2D allocates more time resources to data flows by spatial reuse, and consequently reduces the transmission power of each flow while achieving the same or higher throughput. Similarly, the work \cite{EE-D2D2020} proposed a D2D-enabled multicast scheduling to minimize energy consumption in mmWave cellular networks. To enable cost-effective and flexible heterogeneous cellular networks (HCNs), the power consumption in mmWave backhauling of densely deployed small cells was investigated in \cite{Green}, while the work \cite{Jingli} further extended the results to FD communication scenarios. Furthermore, the authors of \cite{cnA4} performed energy-spectral-efficiency analysis and optimization for HCNs, while the work \cite{cnA5} carried out mobile-traffic-aware energy and spectrum efficiency optimization for large-scale D2D-enabled cellular networks.

Research also paid attention to {\color{black}reducing} the excessive energy waste in IoT nodes, service terminals, or infrastructures. For instance, the high power consumption of radio frequency front-end (RFFE) is a salient and serious issue for mmWave-based mobile devices. To address this problem, the work \cite{Power} enabled discontinuous reception (DRX) to maintain both EE and link reliability, which can be applied to other mmWave terminals and even THz wireless systems. At the network level, the work \cite{Global} proposed a heuristic embedding algorithm with better coordination between the power-aware nodes and link mapping phases, which exhibits superior EE performance. {\color{black}Although this approach has considered realistic factors, including different baseline power consumption of physical nodes and a variety of network equipment at each node, how to extend it to the power-aware migration scenario with flexibility still needs further exploring.} The work \cite{mobility-2019a} proposed the concept of Jamcloud, a system to collect and aggregate the computation capacities of congested vehicles in the city. By outsourcing the BS's baseband signal processing to a nearby vehicular cloudlet, rather than the remote cloud center, substantial energy can be harvested from the jammed vehicles, which would otherwise be unused or wasted.
\revise{Table~\ref{table-EE} summarizes some key schemes of energy efficiency optimization for mobile mmWave applications.}

In recent years, researchers have shown that the mobility of users affects the energy consumption of devices and have conducted some targeted evaluations. For example, the authors of \cite{DTN} compared resource consumption of Epidemic, PRoPHET, and Spray-and-wait protocols under different mobility models, and especially observed remaining energy, delivery probability, and overhead ratio performance. In the future, it is believed that EE can be further improved based on the statistics features extracted from mobility behaviors and big data analysis on mobility patterns \cite{mobility-2016,mobility-2019}.

\begin{figure}[!t]
\begin{center}
\includegraphics[width=\columnwidth]{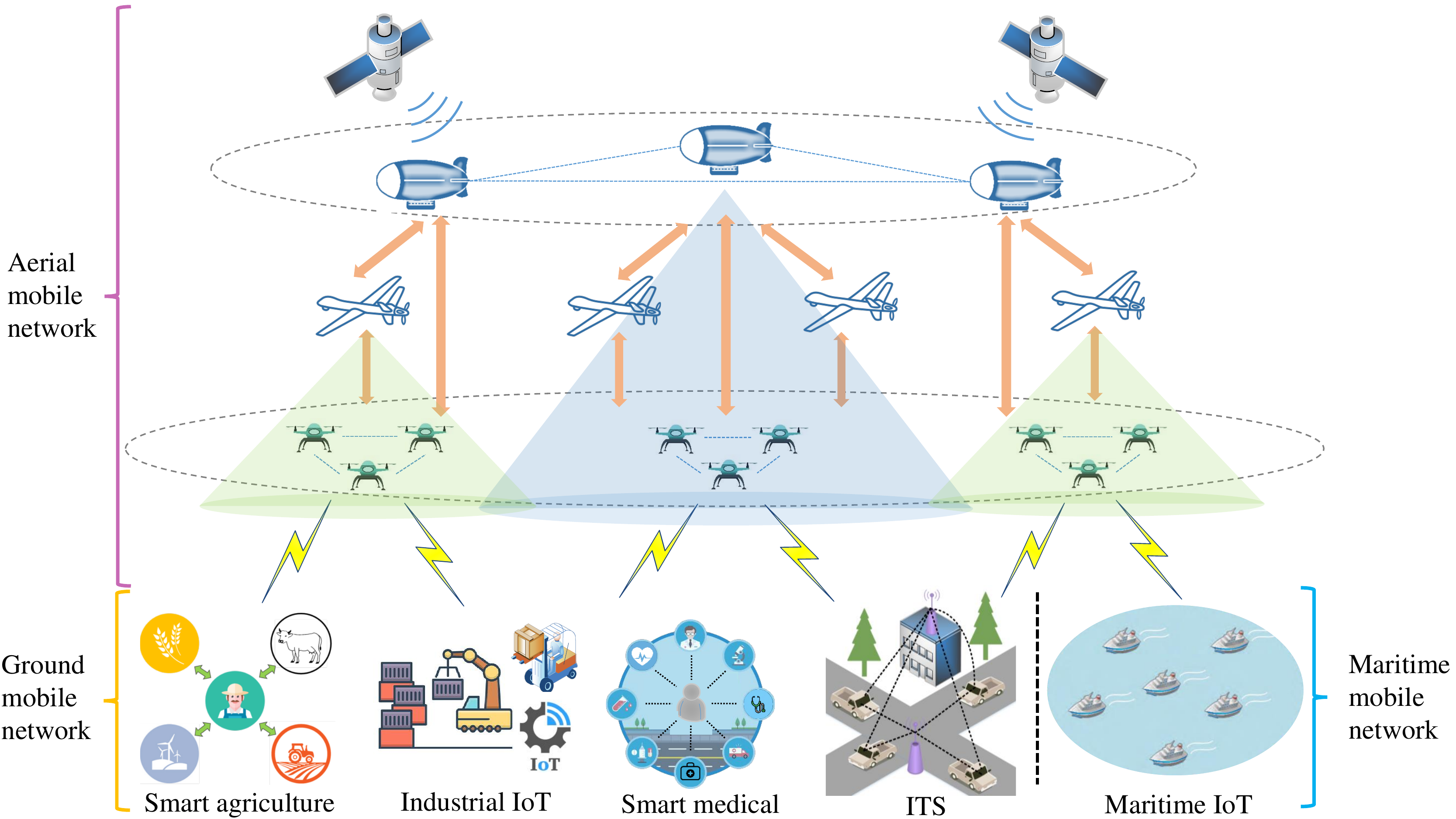}
\end{center}
\vspace*{-2mm}
\caption{The mobile communication system towards future.}
\label{fig:3D}  
\vspace*{-5mm}
\end{figure}

\section{Future Research and Open Issues}\label{S5}

Although the mobility support technologies for mmWave systems have been widely researched in the past years, there are specific issues over the future communications owing to the newly proposed requirements and upcoming applications. We list several open directions that deserve further investigation.

\subsection{\revise{MmWave Enabling HetNets}}\label{S5-1}

As illustrated in Fig.~\ref{fig:3D}, future mobile connections will be extended to aerial and maritime, supporting aerial to aerial (A2A), aerial to ground (A2G), aerial to sea (A2S), sea to sea (S2S), sea to ground (S2G), and ground to ground (G2G) communications. Essentially, the future global SAGS mobile network is composed of huge variety of HetNets \cite{6G-2019,C1,9086520,SAGE-2021}. MmWave technology plays a key role in mobile communications for 5G and beyond. {\color{black} In particular, mobile mmWave communications are suitable candidates for many of these future component HetNets.} Although mmWave signals are unsuitable for long-range propagation, mmWave communications can coexist with various existing and upcoming HetNets, having a wide range of applications in smart agriculture, intelligent industry, ITS, smart medicine, UAV, and maritime system.  {\color{black} Specifically, various existing mmWave solutions discussed in Section~\ref{S4} can be extended to these future HetNet applications but technical issues will become much more challenging than those outlined in Section~\ref{S4}.} Compared to conventional terrestrial networks, the applications of mmWave technology to these future HetNets impose further requirements for measurement, channel modeling, and new approaches of frequency planning as well as interference management.

\revise{Additionally, mobility management is one of the research hotspots in mmWave enabling HetNets.} In the upcoming era, mobility not only exists among smart terminals, e.g., sensor nodes, smart cellphones, and wearable devices, but also covers aerial access points and various mobile relay nodes. Therefore, mobility modeling is in complex 3D space and should incorporate the characteristics of application scenarios, such as the mobility patterns of aircraft's trajectory and the aerodynamic constraints. Furthermore, several essential factors, e.g., collision avoidance, delay constraint, and handover control, in such complex HetNet architecture have drawn attention \cite{SurveyM,MobMan2020}, and more research efforts are warranted to investigate the effective corresponding solutions.

{\color{black}
\subsection{Network Security}\label{S5-2}

As explained in the previous subsection, mmWave technology offers a promising solution to various future mobile HetNet applications. Different from static communications, mobility introduces more challenging requirements on network security, which mainly includes twofold.

\subsubsection{Efficient Authentications}

To provide reliable services for mmWave networks, {\color{black}message authentication} is an essential technique. However, mobility introduces frequent handover among mmWave small cells/HetNets, and therefore multiple authentications between different small cells/tiers/networks are required, which imposes high communication costs and unnecessary latency \cite{InfSecAccess}. Accordingly, authentication schemes are required to be more efficient and several representative methods have been developed. Duan et al. \cite{ICCSDN} proposed a software defined networking-enabled (SDN-enabled) fast authentication method by leveraging weighted security context transfer, to realize the increase of authentication accuracy and the decrease of latency.
Cooperative message authentication for a V2X network was proposed and analyzed in \cite{SXM}, where authentication units are the fleet rather than vehicles, thus reducing authentication messages and saving communication resource. In the future, bringing intelligence and programmability into further optimization of handover authentication will be useful for both attack defence and EE enhancement.

\subsubsection{Flexible Security Mechanisms}

Since the mobility speeds and computation capability of mobile devices in mmWave networks differ greatly, flexible security mechanisms are necessary. For example, mobile sensor nodes are power constrained, thus acquiring for energy efficient and lightweight security algorithms in {\color{black}their} microcontrollers, while for high-speed services like self-driving automobile, efficient security techniques of ultra-reliability and low latency are vitally important. Therefore, as one of the key approaches to ensure security in various mmWave networks, flexible security mechanisms need to be further developed in the future.
}

For individual mmWave links, physical layer security (PLS) approaches offer effective means of secure communications \cite{PLS2019,PLS2019a,InfSec2021}. The work \cite{Secure1} exploited a cooperative diversity scheme to achieve a superior secrecy rate in an energy-constrained cognitive radio network (CRN). The studies \cite{Relay6,Secure3} considered aerial mmWave communications, where part of UAVs worked as jammers to eavesdropping channels, to realize a cooperative security mechanism.
As mobile mmWave applications {\color{black}in the future} are diverse, it is necessary to further develop flexible and effective PLS mechanisms.

{\color{black}
\subsection{Performance Optimization}\label{S5-3}

Two new directions for performance optimization in mmWave mobile networks are highly desirable. The first one is to achieve performance improvement by enhancing hardware efficiency, while the second one is to conduct resource management dynamically for performance enhancement.

\subsubsection{Hardware-algorithm Co-design}

Due to the short wavelength of mmWave signals, a large antenna array can be packed into a small physical dimension to enable the deployment of massive MIMO systems for mobile devices \cite{HBF2019a}. Massive MIMO technique helps to solve the spectrum congestion problem and support high data rate in mobile mmWave network.
{\color{black}In practice,} however, it requires a trade-off between spectral efficiency and hardware efficiency, because power consumption and hardware complexity of beamformer are proportional to the number of phase shifters.
Therefore, it is highly desired to achieve {\color{black}a} cost-effective mmWave transceiver solution by designing hardware-efficient hybrid precoding/beamforming architecture. Inspired by this idea, the work \cite{JSTSP2018} proposed a hybrid precoding method{\color{black}, which provided} a flexible way to trade off spectral efficiency with hardware complexity.
}

\subsubsection{Dynamic Resource Management}

Mobility introduces high Doppler spreads and frequent handover among small cells in the mmWave networks, which decrease the system capacity and energy efficiency. To cope with these problems, highly effective dynamic resource management mechanisms are needed for mobile mmWave systems \cite{ResMan2020}.

Mobile mmWave terminals have limited battery life and computational capacity, which imposes a significant difficulty for supporting the ever-growingly intensive computation demands in the mobile process. Emerging offloading and content-aware caching offer effective methods for battery saving, which in turn, however, may introduce enormous network traffic and significant communication delay, harmful to delay-sensitive critical applications including automatic driving, IIoT, railway control, etc. Therefore, dynamic resource management is essential to achieve a balanced solution to this challenge for future intelligent mobile mmWave terminals. 
The promising rate-aware smart resource scheduling for mobile devices can compensate for adverse effects caused by frequent switching and save power.

From the system perspective, the evolution of classic self-organizing networks to intelligent self-organizing networks is called for, in order to cope with the mobility-caused dynamic routing, complex interference, coverage problem, and capacity maintenance. Therefore, big data analysis functioning and {\color{black}artificial intelligence (AI) enabled controllers} are currently integrated into networks to dynamically and intelligently manage resources.
\revise{As an example, by finding the best user association with Q-learning, the work \cite{UASurvey} significantly decreased handover frequency in mmWave networks with dense small cells.}
Besides, to overcome the spectrum crunch, cognitive spectrum sharing will also be an essential part of network resources management and optimization.

\subsection{AI \revise{Integration}}\label{S5-4}

{\color{black}
Another future research direction is utilizing AI techniques in mmWave mobile communications. Typically, mmWave bands are considered as the de-facto candidate for the Gbps transmission demand to support future mobile applications like self-driving automobile in V2X, real-time sensing in IIoT, and online videos of mobile users. 
However, mobility introduces several new challenges in the system, which can no longer be well solved by traditional methodologies.

\subsubsection{Beam Alignment Maintenance}

Dynamic conditions make it challenging to maintain beam alignments between mobile devices since frequent and adaptive beam alignments are required \cite{BeamINF}.
{\color{black}{This is a critical issue for mmWave, THz and free-space optical (FSO) communications that use directional communications with sharp beams, but only for mmWave systems the hardware components and beam management algorithms have been progressed sufficiently to be leveraged on commercial platforms. In particular, learning-based algorithms offer a flexible solution.} For example, {\color{black}the work \cite{V2XINF} incorporated FML} in a dynamic mmWave vehicular network to conduct beam selection. Facilitated with the ability of autonomous exploration, learning, and adaptability to the environment, mobile BSs/UEs can conduct optimal beam selection, which maximizes the overall network capacity \cite{V2XINF}. Likewise, Satyanarayana et al. \cite{BeamTVT} designed a deep learning-aided beam-alignment scheme in a mmWave vehicular scenario, to achieve the target performance with lower complexity. 
{Ma {et al.} \cite{BeamNew} developed a deep learning-assisted calibrated beam training approach, which achieved significantly higher beamforming gain with smaller beam training overhead compared with the conventional and existing deep-learning based counterparts. This scheme was also capable of handling mobile scenarios.}
Generally speaking, research in this field is still limited and mainly concentrates on V2X communications. How to enable intelligent beam alignments in high speed mobility requires further efforts.

\begin{figure}[!tp]
\begin{center}
\includegraphics[width=\columnwidth]{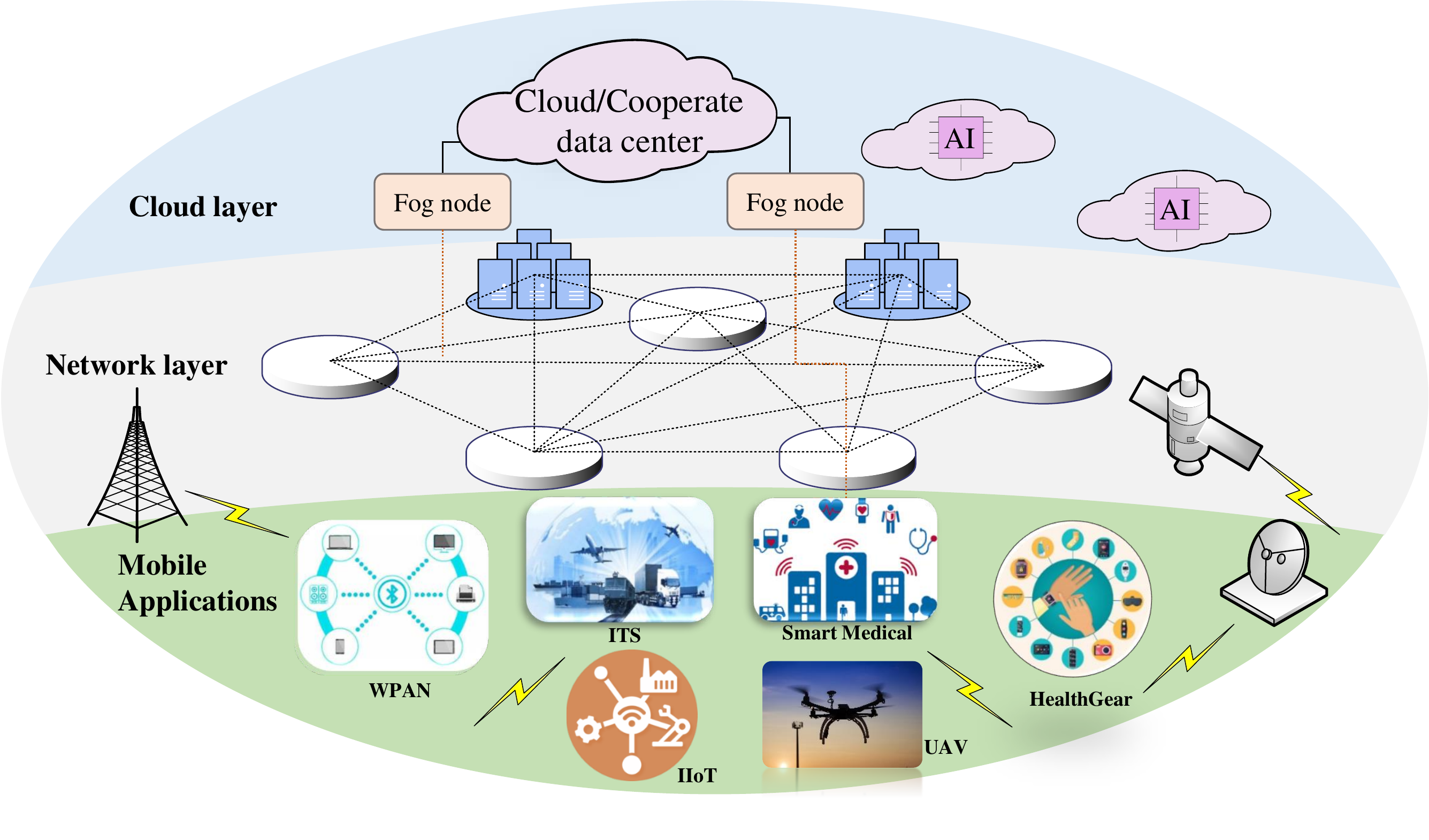}
\end{center}
\vspace*{-4mm}
\caption{AI-enabled network architecture for 5G and beyond.}
\label{fig:Network2} 
\vspace*{-4mm}
\end{figure}

{\color{black}
\subsubsection{AI-enabled Network}

The future global SAGS network is highly dynamic and extremely complex due to its scale, density and diverse mobility scenarios, where traditional optimization approaches are no longer capable of achieving system optimization. A recent new trend is to optimize mmWave systems with AI techniques \cite{AIWCX}. For mobile applications like ITS, although mmWave communications enable Gbps transmission rates for the sensor data, traditional edge nodes of limited power and computational capability are incapable of achieving massive content delivery and data fusion.

A practical solution is to enable intelligent mobile computing, i.e., collaborating the capacity of cloud with edge nodes to handle the requirements of devices adaptively, which can fully utilize the processing capability of the overall system. Therefore, future mobile networks shown in Fig.~\ref{fig:Network2} are incorporated with cloud-edge-collaborated AI for better services. There are three layers, including mobile applications layer, network layer and cloud layer in the architecture, where mmWave technique enables efficient transmission of application data, while intelligent nodes accomplish better training of learning model and cooperated data center prompts smart data fusion to realize dynamic system-level optimization. As an advantage, this  cloud-edge-collaborated AI architecture fully {\color{black}utilizes} the virtualization and flexibility for mobile network layering, and therefore it offers numerous new network services according to different network functionalities and mobility patterns \cite{2020network}. For example, to save power, the recently proposed deep neural networks (DNN) partitioning technology \cite{AIC} optimizes the computation offloading between the mobile devices and the cloud, to provide opportunities for system-level optimization. Additionally, to handle data efficiently while also preventing key information from leakage, common data with extensive computational requirements can be uploaded to the cloud with {\color{black}the mmWave technique} for model training but the sensitive information {\color{black}is} kept at the mobile terminals to protect privacy. 

As future mobile systems are highly heterogeneous, extensive further research is warranted to investigate how to make intelligent scheduling among various devices and processing cores, how to process big data efficiently, and how to make accurate mobile predictions with AI \cite{AIf6G2020,AIf6G2020a}.
}

\subsection{Integration of Geographical Information}\label{S5-5}

\revise{In recent years, mmWave based A2G communication has become a promising candidate for critical mobile applications, such as emergency rescue, mobility prediction, and transportation planning, due to its high reliability, excellent flexibility and large bandwidth availability \cite{UAV_1}. In these critical applications, the geographical information is crucial.}
First of all, geographical information is an important part of interaction messages among terminals and aerial networks, which is transmitted through aerial mobility management cores \revise{with mmWave technique} to help reducing unnecessary switches and to achieve an effective update mechanism in aerial networks \cite{IGI3}. For example, in mmWave satellite communication networks, binding the tracking area with geographic location can avoid frequent updates of tracking area caused by moving satellites. 
\revise{Secondly, time-space features together with geographical information pave the way for the development of intelligent cruising algorithms which enable a UAV to fly out of a blockage zone and establish LoS mmWave communications with mobile users, resulting in system performance improvement \cite{UAV_1}.}
Last but not least, mmWave provides Gbps transmission {\color{black}rates} for the aerial devices to collect and transmit the massive trajectory data containing geographical information of mobile terminals. By analyzing these trajectory data, the mobility pattern of users can be revealed, paving the way for ITS construction.

\revise{In summary, there are} three applications of integrating geographical information in future mmWave HetNets. Firstly, when executing measurement, cell selection or reselection, the real-time positions of satellites/aerial nodes captured at terminals can be introduced as a specific condition for triggering measurement reports or as an assistant for the decision at the network side \cite{IGI4}. Another promising perspective hopes to reveal the correlation-relationship between user preference, mobility regularity, social connection, and time-space feature \cite{IGI2018}.  Finally, \revise{relying on mmWave transmission and} together with big data analysis, it is feasible to reveal the dynamic nature of modern cities \cite{mobility-2016,mobility-2019}. 

\subsection{\revise{Smart and Controllable Communication Environment}}\label{S5-6}

Some researchers improve the performance of mmWave mobile networks not by the existing wireless link adaptation techniques at the TX/RX but from modifying the wireless channel between them.
Such technology provides a new degree of freedom to performance enhancement, which {\color{black}is} realized mainly through two approaches \cite{RIS275}.

\subsubsection{Deployment of Large Intelligent Surface}

Large intelligent surface (LIS) enables the communication environment to become intelligent and controllable, and it offers a new approach for transmission improvement in mmWave and THz frequency bands \cite{RIS275,DLIS1}. Its applications in future mobile systems can be considered from three aspects.
\begin{itemize}
\item Coverage expansion: As mmWave and THz frequency signals exhibit inferior transmission and diffraction capabilities compared with their microwave counterparts, the communication links in B5G and 6G era are subject to obstructions. By providing forwarded signal beams between TX and RX, LIS technology optimizes the wireless propagation, expands the coverage, and provides continuous service \cite{DLIS1,DLIS01}.
\item Integration with MIMO: Due to the ever-increasing mobile data traffic, traditional MIMO technology can no longer meet future traffic requirements. By combining LIS with massive MIMO, enormous spatial multiplexing gains can be achieved \cite{DLIS00,LIS2020}.
\item Flexible deployment: LIS technology can be incorporated with the existing access points or infrastructures flexibly to provide ubiquitous access for mobile terminals, emerging as an important part of future intelligent networks \cite{RIS275,DLIS1}.
\end{itemize}

However, the properties of LIS-based systems have not been fully grasped, and further efforts are still needed in the future dynamic LIS design. In particular, three fundamental physical-layer challenges, namely, CSI acquisition, passive information transfer, and low-complexity robust system design, have to be tackled in order to incorporate LIS fully into future mmWave HetNets. Other promising research directions of LIS include edge intelligence and physical-layer security.

\subsubsection{Deployment of Media-Based Modulation}

An alternative to LIS for making the communication environment intelligent and controllable is media-based modulation (MBM) \cite{MBM2019a}. While LIS enhances wireless transmission by optimizing the wireless propagation to expand the coverage and increase the reliability, MBM performs the transmission of information by altering the far-field radiation pattern of reconfigurable antennas through adjusting the on/off status of its available RF mirrors. This creates a completely new dimension of encoding information bits, namely, wireless channel fade realizations themselves through the unique signature of received signals can be utilized to convey information. MBM is one of the newest and the most prominent members of the index modulation family \cite{MBM2019}. A single reconfigurable antenna with $N$ RF mirrors, which only requires a single RF chain, can support the index set of $2^N$ channel realizations, and this index set itself can convey $N$ information bits. Contrasting this with the spatial modulation with $N$ transmit antennas -- its antenna index set can only convey $\log_2(N)$ information bits.

MBM {\color{black}combined} with massive MIMO enables massive machine-type communications for increasing the throughput, supporting a large number of IoT connecting devices and enhancing the detection performance \cite{MBM2020}.
{\color{black}Time-indexed media-based modulation (TI-MBM) \cite{MBM2020a}} is an index modulation scheme where time slots in a transmission frame are indexed to convey additional information bits in MBM. The scheme decides which time slots and RF mirrors in TI-MBM can be activated such that the achievable transmission rate is maximized.

Future research directions include efficient CSI acquisition \cite{MBM2020,MBMct2017} to reduce pilot overhead, reliable reconfigurable antenna design \cite{MBM2019a,MBMrim2020} to fully realize the potential of MBM, and effectively integrating the MBM technology with other promising technologies \cite{MBMfut2021} to optimize the performance of future mobile networks.

\subsection{{\color{black}Evolution to THz}}\label{S5-7}

{\color{black}Driven by the requirements of extremely high data rate and ultra-reliability in emerging applications (e.g., autonomous vehicles, augmented reality, ultra HD video conferencing and streaming), THz-related research has attracted significant attention \cite{THz1, THz2}. 
In particular, Hossan et al. \cite{THz2} studied the feasibility of THz mobile communications in principle. Firstly, with data rates of terabits-per-second (Tbps), it is sufficient to transfer the required data by intermittent connectivity among mobile users. Secondly, it is expected to minimize the impact of the Doppler effect at THz bands, which is crucial for communications in high-speed scenarios.}
{\color{black}In commercial deployments, however, THz mobile communications are facing more unique challenges than mmWave systems. 
First of all, due to high propagation and molecular absorption losses, the communication range of THz bands is further limited, resulting in more frequent handover during mobility \cite{THzAi}. In addition, to design wideband THz transceivers is a major challenge \cite{THzNew}.
Moreover, as large antenna arrays are deployed to overcome the severe path loss, the codebook design for beam switching is computationally complex \cite{THz3}.}

{\color{black}Thus, THz mobile communications are still in its nascent phase, which require the development of innovative solutions on  mobility management, device design and beam tracking. Moreover, it is a promising direction to integrate THz communications with mmWave and sub-6 GHz bands, which provides many opportunities for realistic universal coverage and mobility support \cite{THzFellow}.}

\section{Conclusions}\label{S6} 

The mmWave communication technology as an effective way to support huge mobile data traffic is developing rapidly in mobile networks, especially {\color{black}in the 5G and 6G eras.} Therefore, we have presented a survey on the challenges and opportunities for mobility-aware mmWave communications. This paper can be understood from three parts, including mobility investigation, existing research, and future outlook. Firstly, we have summarized the mmWave applications in mobile scenarios and then present different mobility models to characterize various mobility patterns in different mmWave applications. Secondly, we have introduced the challenges of adopting mmWave systems and have surveyed the potential solutions to the key problems, including channel measurements and modeling, channel estimation, anti-blockage, and capacity enhancement. Finally, we have proposed the open research issues that have not been fully considered to conclude this paper. We hope that the discussions presented in this paper will serve as the reference and provide the guidelines for the researchers pursuing network planning and optimization for mobile mmWave communications.

\bibliographystyle{IEEEtran}


\end{document}